\documentclass{IEEEtran}
\usepackage{cite}
\usepackage{amsmath,amssymb,amsfonts}
\usepackage{multirow}
\usepackage{algorithmic}
\usepackage{graphicx}
\usepackage{textcomp}
\usepackage{subfigure}
\usepackage{color}
\def\BibTeX{{\rm B\kern-.05em{\sc i\kern-.025em b}\kern-.08em
    T\kern-.1667em\lower.7ex\hbox{E}\kern-.125emX}}
\begin{document}
\title{Parallel-Plate Waveguides Formed by Penetrable Metasurfaces}
\author{X.~Ma, M.~S.~Mirmoosa, \IEEEmembership{Member, IEEE}, and S.~A.~Tretyakov, \IEEEmembership{Fellow, IEEE}
%\thanks{This work has been partly supported by the Aeronautical Science Foundation of China (No.~20161853018).}

\thanks{X.~Ma is with the Department of Electronics and Information, Northwestern Polytechnical University, 710129 Xi'an, China, and the Department of Electronics and Nanoengineering, Aalto University, P.O.~Box 15500, FI-00076 Aalto, Finland (e-mail: maxin1105@nwpu.edu.cn).}
\thanks{M.~S.~Mirmoosa and S.~A.~Tretyakov are with 
the Department of Electronics and Nanoengineering, Aalto University, P.O.~Box 15500, FI-00076 Aalto, Finland (e-mail: mohammad.mirmoosa@aalto.fi and sergei.tretyakov@aalto.fi).}}

\maketitle

\begin{abstract}
In this paper, we introduce and study parallel-plate waveguides formed by two penetrable metasurfaces having arbitrary isotropic sheet impedances. We investigate guided modes of this structure and derive the corresponding dispersion relations. The conditions under which transverse magnetic and transverse electric modes can exist are discussed, and different scenarios including lossless (reactive) metasurfaces, gain-and-loss sheets and extreme cases are under general consideration. Resonant and non-resonant dispersive sheets and corresponding extreme cases are investigated. Our theoretical results are confirmed with full-wave simulated results considering practical realizations of the proposed parallel-plate waveguide which exploit two frequency-selective surfaces. Finally, the obtained theoretical formulas are applied to study the dispersion diagrams for waves along resonant sheets at different distances between the two identical or different metasurfaces. We hope that this study is useful for future applications at both microwave and optical regimes.
\end{abstract}


\section{Introduction}
\label{sec:introduction}
\IEEEPARstart{W}{aveguides} serve as enabling components in microwave and optical technologies, transferring electromagnetic energy and signals over a distance~\cite{WGuidebook1,WGuidebook2}. Recently, following the development of artificial impedance surfaces (or metasurfaces)~\cite{FSS1,Holloway,tretyakov,bian_metasurfaces}, novel waveguide structures that exploit impedance surfaces/metasurfaces have been proposed for guiding and control of waves~\cite{sievenpiper1,kabakian,sievenpiper2,bilow,grbic,sievenpiper3,maci1,maci2,engheta,Li,sun,MW35,grbic2}. Such open waveguides, which were initially considered in the previous century for the purpose of radiating the electromagnetic energy~\cite{hessel,hwang}, are supporting propagating surface waves if the structure is uniform along the propagation direction. Surface waves of  arbitrary polarizations can be excited for example by an antenna (emitter) near the surface. 

Surface waves along  impedance boundaries are well understood, as well as waves in closed waveguides with impedance walls. However, there is not enough clear view on what happens if we bring two thin penetrable sheets close to each other and form a waveguide this way. In the limit of zero or infinite sheet impedance we come to the trivial special cases of parallel-plate waveguides bounded by perfect electric conductor (PEC) or perfect magnetic conductor (PMC) boundaries. Here we are interested in the general case when the two sheets are penetrable and can  be characterized by dispersive and resonant sheet impedances. The goal of this paper is to study such waveguides and derive the general  dispersion relations and find the field distributions corresponding to different allowed polarizations. From one side, this kind of waveguide is similar to the parallel-plate waveguide which is consisting of two metal plates (the energy is fully confined inside the waveguide). However, from another side, it is also similar to the classical dielectric slab waveguide which is based on the total internal reflection. Notice that the impedance of the sheets is arbitrary and can be lossy and active as well. 

The paper is organized as follows: In Section~\ref{sec:sus}, we introduce our novel structure operating as a waveguide, and in Sections~\ref{sec:tmp} and \ref{sec:tep}, the dispersion relation is derived for transverse magnetic (TM) and transverse electric (TE) polarizations, respectively. Section~\ref{sec:NonResonant} shows simulated results which correspond to a practical realization based on patch or grid inclusions, and confirms our theoretical results regarding dispersion relations. Section~\ref{sec:resonant} discusses and illustrates different resonant scenarios for symmetric and asymmetric cases. Finally, Section~\ref{sec:con} concludes the paper.                   

%%%%%%%%%%%%%%%%%%%%%%%%%%%%%%%%%%%%%%%%%%%%%%%%%%%
%%%%%%%%%%%%%%%%%%%%%%%%%%%%%%%%%%%%%%%%%%%%%%%%%%%
\section{Structure Under Study}
\label{sec:sus}

\begin{figure}[!t]
	\centerline{\includegraphics[width=\columnwidth]{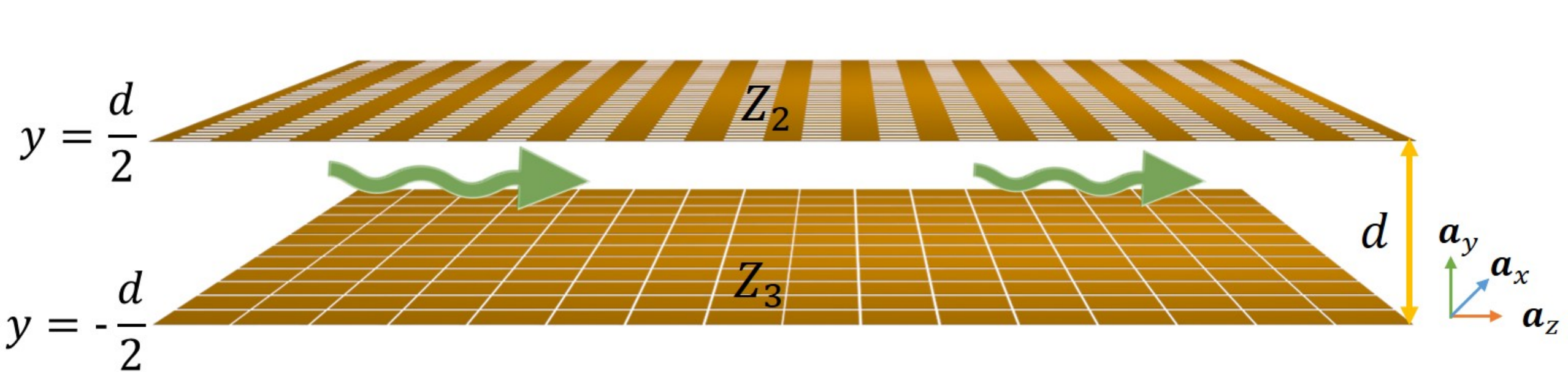}}
	\caption{The structure under study: Two parallel impedance sheets. The wave is propagating along the $z$-direction.}
	\label{fig:st}
\end{figure}
The proposed and studied parallel-plate waveguide is shown in Fig.~\ref{fig:st}. It consists of two parallel penetrable metasurfaces separated by the distance $d$. One sheet is  positioned at $y=d/2$ and consequently the other one is placed at $y=-d/2$. The space between the two sheets is filled by air (vacuum). The sheet impedances of the two metasurfaces are denoted as $Z_2$ and $Z_3$, respectively, and we assume that these values do not depend on the spatial coordinates. We consider scalar-valued impedances assuming that the surfaces are either isotropic or we mean the value which is meaningful for a specific polarization which we discuss. Notice that, generally speaking, both impedances can be complex-valued with non-zero resistive and reactive parts. We orient the $z$-axis along the wave propagation direction.

Under the above assumptions, there is a possibility to decompose the fields into two polarizations: The transverse magnetic polarization (TM waves) and the transverse electric polarization (TE waves). The polarization is defined with respect to the propagation direction. According to Fig.~\ref{fig:st}, the $xy$-plane is the transverse plane and the propagation direction (the longitudinal direction)  corresponds to the $z$-axis. TM waves have a longitudinal electric-field component ($E_z\neq0$) and TE waves have a longitudinal magnetic-field component ($H_z\neq0$). For each polarization, we derive the dispersion relation and investigate different scenarios that can be realized for various sheet impedances. 

Note that in this study, the medium between the sheets is assumed to be air, as mentioned earlier. In the following sections, we show that such assumption prominently affects the wave vector (specifically, the real and imaginary parts of the transverse component of the wave vector) between the sheets. However, in most practical realizations, there is a substrate between the sheets whose relative permittivity can be different from unity. From this point of view, it will be interesting to investigate how controlling the relative permittivity of the substrate may result in controlling the field distribution and, generally, the propagation of guided waves. We hope to report on this investigation in near future. In addition to the above assumption about the substrate material, here we consider only electric sheets. In other words, the sheets support only electric surface currents. Using duality, it is possible to find the corresponding solutions for the case of magnetic sheets supporting magnetic surface currents.  In future, it can be also interesting to investigate waves along  magneto-electric sheets supporting both electric and magnetic currents and exhibiting bianisotropy.  

%%%%%%%%%%%%%%%%%%%%%%%%%%%%%%%%%%%%%%%%%%%%%%%%%%%%%%%%%%%%%%%%%%%%%%%%%%%%%%%%%%%%%%%%%%%%%%%%%%%%%%%%%%%%%%%%%%%%%%%%%%%%%%%%%%%%%%%%%%%%%%%%%%%%%%%%%%%%%%%%%%%%%%%%%%%%%%%%%%%%%%%%%%%%%%%%%%%%%%%%%%%%%%%%%%%%%%%%%%%%%%%%%%

\section{Transverse Magnetic Polarization}
\label{sec:tmp}
In this section, we study guided modes of TM polarization, when the longitudinal component of the magnetic field vanishes. Solving the Helmholtz equation, we can write for modes bound to the waveguide
\begin{equation}
E_z=\Big[E_{\rm{e}}\cos(hy)+E_{\rm{o}}\sin(hy)\Big]e^{-j\beta z},
\end{equation}
for the field inside the waveguide and 
\begin{equation}
\begin{split}
&E_z=Ae^{-\alpha(y-{\frac{d}{2}})}e^{-j\beta z},\,\,\,\,y\geq{\frac{d}{2}},\cr
&E_z=Be^{\alpha(y+{\frac{d}{2}})}e^{-j\beta z},\,\,\,\,y\leq-{\frac{d}{2}},
\end{split}
\end{equation}
for the outside space. Parameters $E_{\rm{e}}$, $E_{\rm{o}}$, $A$ and $B$ are unknown coefficients which must be found later. Here, $\beta$ is the phase constant and $\alpha$ corresponds to the attenuation of the field outside of the waveguide. Since the material inside and outside the waveguide is assumed to be the same (free space), we conclude that
\begin{equation}
h=-j\alpha.
\end{equation} 
This is important because it shows that the transverse component of the wave vector inside the waveguide is also purely  imaginary. As we will see later, this property restricts the number
of modes to four, in contrast to closed waveguides which can support infinitely many propagating modes.
%It results in changing the functions $\sin(jx)$ and $\cos(jx)$ into $j\sinh(x)$ and $\cosh(x)$, respectively (here, $x$ is assumed to be a real variable). 
Expressing the other components of the fields via the longitudinal component, we have
\begin{equation}
\begin{split}
&E_y\vert_{-{d/2}<y<{d/2}}={j\beta\over h}\Big[E_{\rm{e}}\sin(hy)-E_{\rm{o}}\cos(hy)\Big]e^{-j\beta z},\cr
&H_x\vert_{-{d/2}<y<{d/2}}={j\omega\epsilon_0\over h}\Big[-E_{\rm{e}}\sin(hy)+E_{\rm{o}}\cos(hy)\Big]e^{-j\beta z},
\end{split}
\end{equation} 
and
\begin{equation}
\begin{split}
&E_y\vert_{y\geq{d/2}}=-j{\beta\over\alpha}Ae^{-\alpha(y-{d\over2})}e^{-j\beta z},\cr
&H_x\vert_{y\geq{d/2}}=j{\omega\epsilon_0\over\alpha}Ae^{-\alpha(y-{d\over2})}e^{-j\beta z},\cr
&E_y\vert_{y\leq-{d/2}}=j{\beta\over\alpha}Be^{\alpha(y+{d\over2})}e^{-j\beta z},\cr
&H_x\vert_{y\leq-{d/2}}=-j{\omega\epsilon_0\over\alpha}Be^{\alpha(y+{d\over2})}e^{-j\beta z}.
\end{split}
\end{equation} 
Knowing that the electric field (its tangential component) is continuous at the surfaces $y=d/2$ and $y=-d/2$, we can find the unknown coefficients for the amplitudes $A$ and $B$. However, for finding the dispersion relation we use the boundary condition related to the discontinuity of the magnetic field:
\begin{equation}
Z_i\Big[\mathbf{a}_{\rm{n}}\times(\mathbf{H_1}-\mathbf{H_2})\Big]=\mathbf{E_{\rm{t}}},
\label{eq:bound}
\end{equation}  
in which $Z_i=Z_2$ is the sheet at $y=d/2$, and $Z_i=Z_3$ denotes the sheet impedance at $y=-d/2$. In the above equation $\mathbf{a}_{\rm{n}}$ is the normal unit vector to the sheet plane and $\mathbf{E}_{\rm{t}}$  represents the tangential component of the electric field which is continuous across the impedance sheets. Here, we assume that the sheets are supporting only electric surface currents.   
After some algebraic manipulation, we find that
\begin{equation}
\begin{split}
&\begin{bmatrix}
\cosh(\alpha d/2)+j{Z_2\omega\epsilon_0\over\alpha}e^{\alpha d/2} & {Z_2\omega\epsilon_0\over\alpha}e^{\alpha d/2}-j\sinh(\alpha d/2)\\\\
\cosh(\alpha d/2)+j{Z_3\omega\epsilon_0\over\alpha}e^{\alpha d/2} & {-Z_3\omega\epsilon_0\over\alpha}e^{\alpha d/2}+j\sinh(\alpha d/2)
\end{bmatrix}\cr
&\cdot\begin{bmatrix}
E_{\rm{e}}\\\\
E_{\rm{o}}  
\end{bmatrix}
=
0.
\end{split}
\end{equation}
Equating the determinant of the square matrix to zero, we find the dispersion relation:
\begin{equation}
\alpha^2\Big(e^{-2\alpha d}-1\Big)=j2\omega\epsilon_0\Big(Z_2+Z_3\Big)\alpha-4\omega^2\epsilon_0^2Z_2Z_3.
\label{eq:gedisrelz3z2} 
\end{equation}
The left-hand side of the above equation is purely real while the right-hand side is complex-valued. Therefore, the imaginary part of the right-hand side should vanish. Let us assume that $Z_2=R_2+jX_2$ and $Z_3=R_3+jX_3$. It is easy to see that there are two cases in which this imaginary part vanishes. The first case corresponds to reactive sheets where $R_2=R_3=0$. In this case, the dispersion relation can be simplified as
\begin{equation}
\alpha^2\Big(e^{-2\alpha d}-1\Big)+2\omega\epsilon_0\Big(X_2+X_3\Big)\alpha-4\omega^2\epsilon_0^2X_2X_3=0.
\label{eq:anonloss}
\end{equation}
The second case is when the structure has balanced gain and loss. In other words, one sheet is resistive and the other one in active such that $R_2=-R_3$. However, this requirement is not sufficient for having real positive solutions for the attenuation constant. In addition, the reactive  components of the sheet impedance must be also equal, i.e. $X_2=X_3=X$. This means that both sheets must be either inductive or capacitive. 
We write the dispersion relation for this case as
\begin{equation}
\alpha^2\Big(e^{-2\alpha d}-1\Big)+4\omega\epsilon_0X\alpha-4\omega^2\epsilon_0^2\Big(X^2+R^2\Big)=0.
\label{eq:alossgain}
\end{equation}  
As a check, we see that in the limiting case of PEC boundaries ($R,X\rightarrow 0$) we come to the solution $\alpha=\pm j\pi m/d$, where $m=0,1,2,\dots$. Importantly, while closed waveguides have an infinite (countable) number of modes, as soon as  the walls become penetrable, we end up with only four or two guided-mode solutions. The relation $Z_2=-Z_3^*$ reminds the well-studied parity-time symmetric systems, usually defined in terms of such relation for the permittivity. In the following, we discuss this issue and show that the sheets must be obligatorily inductive in order to allow a guided-mode solution.

%%%%%%%%%%%%%%%%%%%%%%%%%%%%%%%%%%%%%%%%%%
\subsection{Reactive metasurfaces}

Equation~\eqref{eq:anonloss} gives the general dispersion relation. Firstly, let us assume that the sheet reactances are not frequency dispersive.\footnote{Since the sheet reactance of any passive metasurface is always dispersive, the results obtained under this assumption can be used only at a given frequency (calculating the propagation constant at a certain frequency, one should substitute the reactances of the sheets at this frequency).}  Also, let us write $X_3=nX_2$ in which $n$ can be positive or negative. Equation~\eqref{eq:anonloss} simplifies as
\begin{equation}
\omega={\alpha\over4\epsilon_0nX_2}\bigg[(n+1)\pm\sqrt{(n+1)^2-4n(1-e^{-2\alpha d})}\bigg].
\label{eq:omegawTM}
\end{equation} 
As is clear from the above expression for the angular frequency, we can distinguish four different scenarios. Two scenarios are associated with $X_2>0$ while $n<0$ or $n>0$, and the other two correspond to the negative $X_2$. It can be shown that when two sheets are inductive ($X_2$ and $n>0$), there are always two TM-polarized modes. However, if one sheet is capacitive and the other one is inductive ($n<0$), the above equation tells that only one TM mode exists which can propagate along the structure and is bounded at the sheets. Finally, when both sheets are capacitive ($X_2<0$ and $n>0$), there is no guided TM mode. These conclusions are in agreement with the fact that a single inductive surface supports TM surface waves and single capacitive surface supports TE surface waves~\cite{Antenna}. In chapter~19 of the Ref.~\cite{Antenna}, there is a discussion about this subject (guided and leaky waves on planar open structures).

%%%%%%%%%%%%%%%%%%%%%%%%%%%%%%%%%%%%%%%%%%
\subsection{Investigation of extreme scenarios}
Here we will discuss two extreme scenarios. The first scenario assumes that the surface impedance of one of the metasurfaces is infinite, which corresponds to surface modes at only one metasurface~\cite{FSS1}. The second scenario assumes that the surface impedance of one metasurface approaches zero, and we deal with an impenetrable reactive boundary realized as a reactive sheet  parallel to a ground plane. 

Let us first investigate guided modes along one penetrable reactive sheet. In the general solution, we require that the impedance of the other sheet is infinite (i.e.~open circuit). From the dispersion relation \eqref{eq:anonloss} for TM polarization, reactance $X_3$ can be expressed as
\begin{equation}
X_3=nX_2={\alpha^2\Big(e^{-2\alpha d}-1\Big)+2\omega\epsilon_0X_2\alpha\over 4\omega^2 \epsilon_0^2X_2-2\omega\epsilon_0\alpha }.
\label{eq:x3TM}
\end{equation}	 
Only when the denominator of \eqref{eq:x3TM} is equal to zero, $X_3$ is infinite for finite values of $X_2$, which results in two solutions for $\omega$:
\begin{equation}
\begin{split}
&\omega\vert_{n\to\infty}=0,\cr
&\omega\vert_{n\to\infty}={\alpha\over 2\epsilon_0 X_2}.
\label{eq:omegaINFTM}
\end{split}
\end{equation}	
Since for surface-bound modes $\alpha>0$, $X_2$ should be obligatorily positive.  If $X_2$ is a  negative reactance, the attenuation constant $\alpha$ should be negative as well to keep $\omega$  positive, which means leaky waves. As expected, a TM surface wave can be support by an inductive surface for single-metasurface waveguide. Note that solution \eqref{eq:omegaINFTM} differs from the solution for an impenetrable reactive boundary \cite{FSS1} only by the factor $1/2$. 

Based on the dispersion relation, the phase velocity of surface waves along single sheets can be obtained:
\begin{equation}
v_p={\omega \over\beta}=c{\eta_0\over \sqrt{4X_2^2+\eta_0^2} },
\end{equation}	
where $\eta_0$ represents the free-space intrinsic impedance ($\eta_0=\sqrt{\mu_0/\epsilon_0})$ and $c$ is the speed of light. The phase velocity of surface waves is slower than velocity of light, and  inversely proportional to $X_2$. The frequency $\omega$ of the second mode goes to zero meaning that this mode which should be bounded by the second sheet ($y=-d/2$) vanishes because of the absence of the second sheet.

The other scenario corresponds to $n$ approaching zero in \eqref{eq:omegawTM}. When $n$ is approaching zero from the positive-value side, the dispersion relations for the two modes can be written as
\begin{equation}
\begin{split}
&\beta\vert_{n\to 0}={\omega\over c},\cr
&\beta\vert_{n\to 0}={\omega\over c} \sqrt{1 +c^2 {4\epsilon_0^2X_2^2 \over (1-e^{-2\alpha d})^2}}.
\label{eq:ZeroTM}
\end{split}
\end{equation}	
From the first equation of \eqref{eq:ZeroTM}, it can be concluded that when $n$ goes to zero, one of modes propagates as a wave in free space.  When $n$ is approaching zero from the negative side, as mentioned previously, there is only one mode whose dispersion relation is given by the second equation of \eqref{eq:ZeroTM}.
\begin{figure}[t!]	
	\centerline{\includegraphics[width=0.3\textwidth]{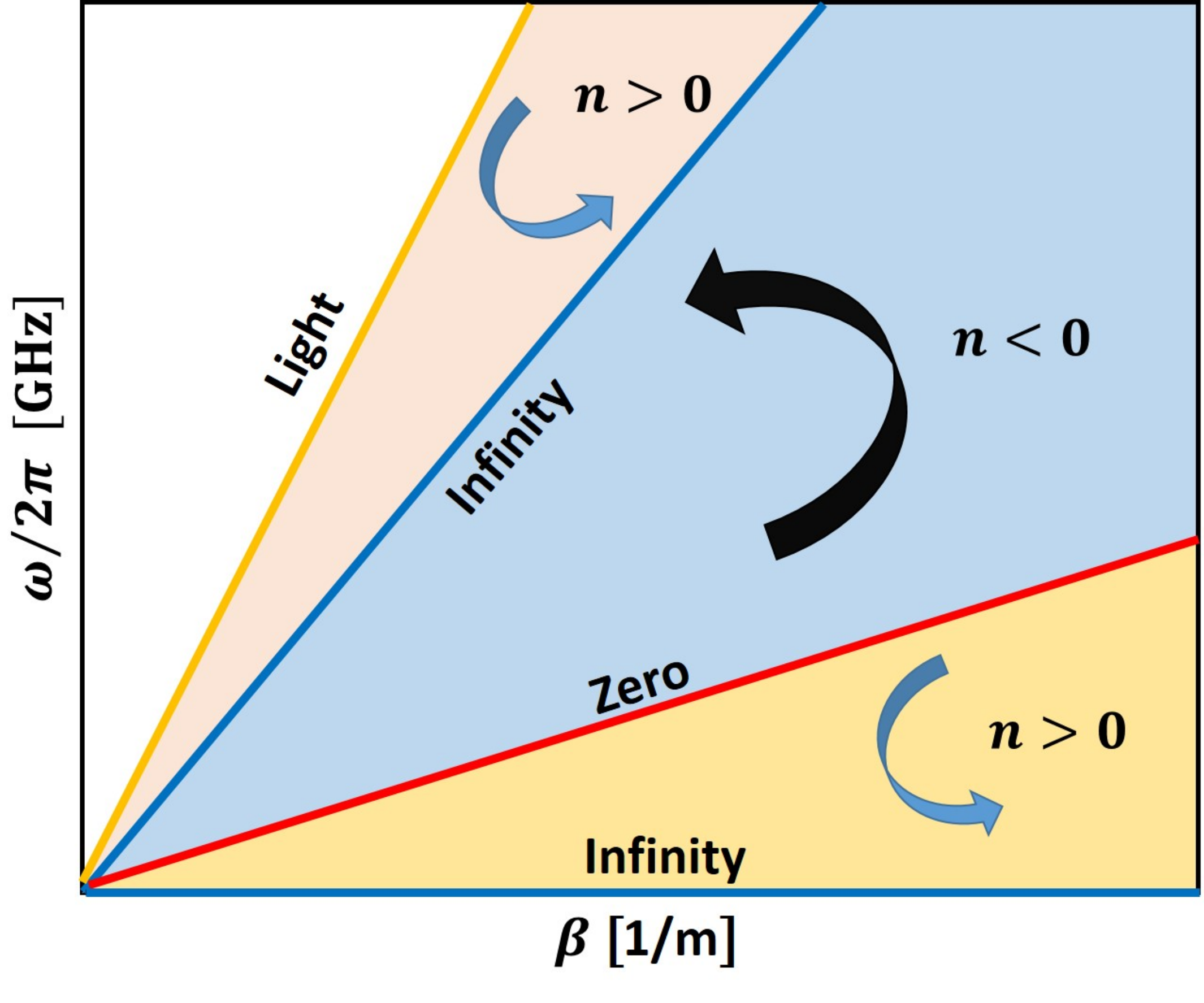}}
	\caption{Distribution of dispersion curves for the TM mode. The yellow line is the light line. Red lines represent dispersion curves for $n\to 0$. The blue lines present the dispersion curves for $n\to\infty$.}
	\label{fig:ExteTM}
\end{figure}

Figure~\ref{fig:ExteTM} shows the dispersion curves for both scenarios. It is clear that as $n$ is approaching infinity, one of the mode curves tends to the zero-frequency line, as shown in blue color. When $n$ is approaching zero, the dispersion curve of one of the modes overlaps with the light line. It is worth to clarify that when $n>0$, the dispersion curves for two modes locate in two separate regions, highlighted in pink and yellow colors in Fig.~\ref{fig:ExteTM}. As $n$ increases, the dispersion curves ``rotate'' in the direction closer to the infinity lines, which are shown in the direction of blue arrows. When $n<0$, the dispersion curves are confined to only one region (the blue region) because only one mode exists in this scenario. As $n$ increases, the dispersion curves ``rotate'' counterclockwise, illustrated by black arrow. According to the second equations of \eqref{eq:omegaINFTM} and \eqref{eq:ZeroTM}, the dispersion curves of the modes for extreme scenarios can be tuned varying the impedance of the sheet ($X_2$) and/or distance ($d$). Therefore, we can arrange the boundaries of three regions (pink and yellow regions for $n>0$ and blue region for $n<0$) by choosing different $X_2$ and $d$.

%%%%%%%%%%%%%%%%%%%%%%%%%%%%%%%%%%%%%%%%%%%%%%%%%%%%%%%%%%%%%%%%%%%%%%%%%%%%%%%%%%%%%%%%%%%%%%%%%%%%%%%%%%%%%%%%%%%%%%%%%%%%%%%%%%%%%%%%%

\section{Transverse Electric Polarization}
\label{sec:tep}
Here, we study guided modes of TE polarization, with zero longitudinal component of the electric field. Similarly to the TM polarization, we can express the longitudinal component of the magnetic field as
\begin{equation}
H_z=\Big[H_{\rm{e}}\cos(hy)+H_{\rm{o}}\sin(hy)\Big]e^{-j\beta z}.
\end{equation} 
The above expression corresponds to the field inside the waveguide. However, the field outside must attenuate, and is written as following: 
\begin{equation}
\begin{split}
&H_z=A e^{-\alpha(y-{d\over2})}e^{-j\beta z},\,\,\,\,y\geq{d\over2},\cr
&H_z=B e^{\alpha(y+{d\over2})}e^{-j\beta z},\,\,\,\,y\leq-{d\over2},
\end{split}
\end{equation}
where $\alpha$ is real and positive. The other components of the fields can be readily obtained from  $H_z$: 
\begin{equation}
\begin{split}
&H_y\vert_{-{d/2}<y<{d/2}}={j\beta\over h}\Big[H_{\rm{e}}\sin(hy)-H_{\rm{o}}\cos(hy)\Big]e^{-j\beta z},\cr
&E_x\vert_{-{d/2}<y<{d/2}}={j\omega\mu_0\over h}\Big[H_{\rm{e}}\sin(hy)-H_{\rm{o}}\cos(hy)\Big]e^{-j\beta z},\cr
\end{split}
\end{equation}
and
\begin{equation}
\begin{split}
&H_y\vert_{y\geq{d/2}}=-{j\beta\over\alpha} A e^{-\alpha(y-{d\over2})}e^{-j\beta z},\cr
&E_x\vert_{y\geq{d/2}}=-{j\omega\mu_0\over\alpha} A e^{-\alpha(y-{d\over2})}e^{-j\beta z},\cr
&H_y\vert_{y\leq-{d/2}}={j\beta\over\alpha} B e^{\alpha(y+{d\over2})}e^{-j\beta z},\cr
&E_x\vert_{y\leq-{d/2}}={j\omega\mu_0\over\alpha} B e^{\alpha(y+{d\over2})}e^{-j\beta z}.
\end{split}
\end{equation}
By imposing continuity of the tangential electric field and discontinuity of the magnetic field (\eqref{eq:bound}) at $y=d/2$ and $y=-d/2$, the unknown coefficients are determined and the dispersion relation can be found from the following characteristic equation:
\begin{equation}
\begin{vmatrix}
j\xi-{\omega\mu_0\over\alpha Z_2}\cosh(\alpha d/2) & -\xi-j{\omega\mu_0\over\alpha Z_2}\sinh(\alpha d/2)\\\\
j\xi-{\omega\mu_0\over\alpha Z_3}\cosh(\alpha d/2) & \xi+j{\omega\mu_0\over\alpha Z_3}\sinh(\alpha d/2)
\end{vmatrix}=0,
\end{equation}
in which $\vert...\vert$ represents the determinant of the corresponding matrix, and $\xi=e^{\alpha d/2}$. After some algebraic manipulations, we find
\begin{equation}
4Z_2 Z_3\alpha^2+j2\omega\mu_0(Z_2+Z_3)\alpha+\omega^2\mu_0^2 (e^{-2\alpha d}-1)=0,
\label{eq:DisR}
\end{equation}
that is rather different from \eqref{eq:gedisrelz3z2}. 
For complex-valued sheet impedances $Z_2=R_2+jX_2$ and $Z_3=R_3+jX_3$, the left-hand side of the above equation is complex-valued. Hence, both the  imaginary and real parts of it must be zero, and we come to the following two equations:
\begin{equation}
\alpha=-{\omega\mu_0\over 2}{R_2+R_3\over R_2X_3+R_3X_2 }
\label{eq:r1}
\end{equation} 
and
\begin{equation}
\begin{split}
4\Big(X_2X_3-R_2R_3\Big)\alpha^2&+2\omega\mu_0\Big(X_2+X_3\Big)\alpha\cr
&+\omega^2\mu_0^2\Big(1-e^{-2\alpha d}\Big)=0.
\end{split}
\label{eq:w2}
\end{equation} 
One can show that if $R_2$ and $R_3$ are not zero, there is only one possibility to have a positive real attenuation constant satisfying both \eqref{eq:r1} and \eqref{eq:w2}, and it is when $R_2=-R_3$. According to \eqref{eq:r1}, we conclude that if $R_2=-R_3$ then the reactive components must be equal, i.e.~$X_2=X_3$, so that both numerator and denominator of \eqref{eq:r1} go to zero. Thus, we again come to the party-time symmetric solution. In contrast to the TM polarization, both metasurfaces must be capacitive in this case of gain and loss. But let us concentrate on the case when the sheets are purely reactive ($R_2=R_3=0$).
Note that all the dispersion relations presented above can be derived using the transverse resonance technique.

%%%%%%%%%%%%%%%%%%%%%%%%%%%%%%%%%%%%%%%%%%%%%%%%%%%%%%%%%%%%%%%%%%%%%
\subsection{Reactive metasurfaces}
Assuming that $Z_2=jX_2$ and $Z_3=jX_3$, \eqref{eq:w2} reduces to
\begin{equation}
4X_2X_3\alpha^2+2\omega\mu_0\Big(X_2+X_3\Big)\alpha+\omega^2\mu_0^2\Big(1-e^{-2\alpha d}\Big)=0.
\label{eq:x1}
\end{equation} 
Let us write $X_3=nX_2$, where $n$ can take any real value. Using this notation, the angular frequencies at which TE modes propagate along the waveguide can be calculated as
\begin{equation}
\omega=-{\alpha X_2\over\mu_0(1-e^{-2\alpha d})}\Big[(1+n)\pm\sqrt{(n-1)^2+4ne^{-2\alpha d}}\Big].
\label{eq:x2}
\end{equation} 
In this equation, the expression under the square root on the right-hand side is always non-negative resulting in two real solutions for the angular frequency. However, only positive solutions are acceptable for $\omega$. According to \eqref{eq:x2}, three possibilities can realize. If $X_2<0$ and $n>0$ -- both sheets are capacitive -- there are always two TE-polarized modes. If one sheet is capacitive and the other one is inductive ($n<0$), only one guided mode exists. Finally, if both sheets are inductive, the structure does not support any TE-polarized guided modes. Recall that for the TM polarization we had similar possibilities. But for that polarization, the two metasurfaces must be both inductive, or one inductive and the other one capacitive. 

As mentioned above, if the two sheets are capacitive, there are two modes. Interestingly, one of those modes suffers from a cut-off frequency at which the attenuation constant $\alpha$ approaches zero. Using \eqref{eq:x2}, it can be shown that
\begin{equation}
f_{\rm{cut-off}}\approx{n+1\over2\pi\mu_0 d}\vert X_2\vert.
\label{eq:cutoff}
\end{equation}

%%%%%%%%%%%%%%%%%%%%%%%%%%%%%%%%%%%%%%%%%%%%%%%%%%%%%%%%%%%%%%%%%%%%%
\subsection{Investigation of extreme scenarios}
Here, similarly to the previous section, we study the two extreme scenarios for the TE polarization. The first and the second scenario correspond to the infinite and zero surface impedance of one of the sheets, respectively. Using the same method as we applied for the TM polarization, we assume that $n$ is approaching infinity in \eqref{eq:x2}. This gives rise to two solutions of $\omega$:
\begin{equation}
\begin{split}
&\omega\vert_{n\to\infty}=\infty,\cr
&\omega\vert_{n\to\infty}=-{2\alpha X_2\over\mu_0 }.
\label{eq:omegaINFTE}
\end{split}
\end{equation}
The first mode  which should be bounded to the second sheet ($y=-d/2$) disappears because of the absence of the second sheet. The second solution is proportional to $\alpha X_2$ which indicates that reactance $X_2$ should be always negative to ensure that $\omega>0$. The explanation is that only capacitive surfaces can support TE surface waves. 

The other scenario is realized when $n$ approaches zero in \eqref{eq:x2}. When $n$ tends to zero from  the positive side, the dispersion relations for the two modes are obtained as following:
\begin{equation}
\begin{split}
&\beta\vert_{n\to 0}=0,\cr
&\beta\vert_{n\to 0}={\omega\over c} \sqrt{1 + c^2{\mu_0^2(1-e^{-2\alpha d})^2\over 4 X_2^2}}.
\label{eq:ZeroTE}
\end{split}
\end{equation}	
It can be found that as $n$ tends to zero, one of the modes disappears, and the other mode can be controlled by the sheet impedance $X_2$ and the distance between the two sheets ($d$). Interestingly, the existing mode suffers from a cut-off where  the attenuation constant $\alpha$ approaches zero. The cut-off frequency can be expressed as
\begin{equation}
f_{\rm{cut-off}}\approx{\vert X_2\vert\over2\pi\mu_0 d}.
\end{equation}
When $n$ is approaching zero from the negative side, there is only one supported mode, and its dispersion relation is given by the second equation of \eqref{eq:ZeroTE}.

\begin{figure}[t!]	
	\centerline{\includegraphics[width=0.3\textwidth]{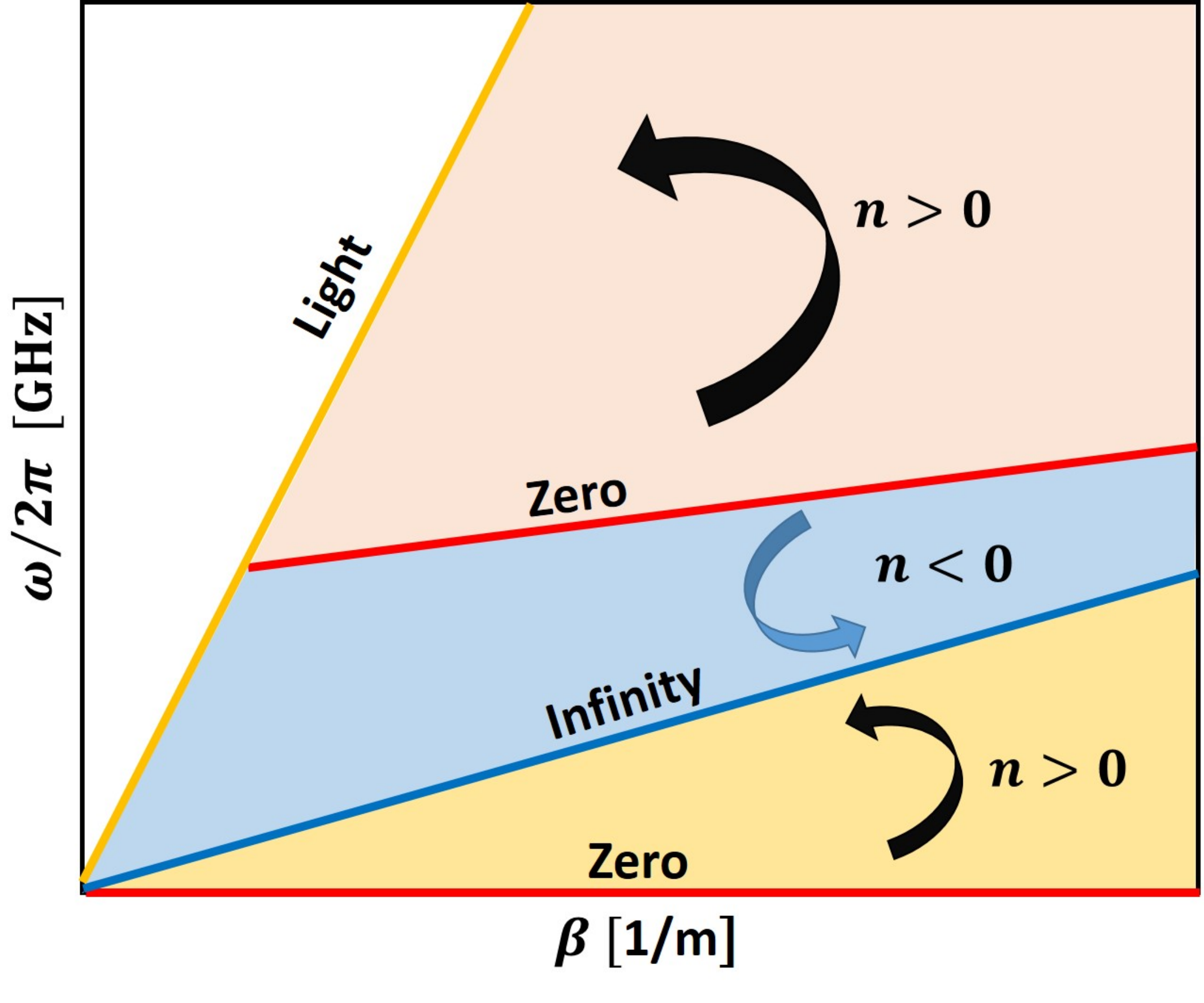}}
	\caption{Distribution of dispersion curves for the TE mode. The yellow line shows the light line. Red lines represent dispersion curves for $n\to 0$. Blue lines show the dispersion curves for $n\to\infty$.}
	\label{fig:ExteTE}
\end{figure}

Figure~\ref{fig:ExteTE} shows the distribution of dispersion curves for the two extreme scenarios for the TE polarization. As it is seen, when $n>0$, the dispersion curves for the two modes locate in two separate regions (pink and yellow regions). As $n$ increases, the dispersion curves move closer to the infinity lines, as shown by black arrows (counterclockwise). Notice that one of modes always has a  cut-off, approximately given by \eqref{eq:cutoff}. Recall that for the TM polarization, both modes have no cut-off. When $n<0$, all dispersion curves locate in one region (the blue region), because only one mode exists. As $n$ increases, the dispersion curves ``rotate'' in the direction shown by the blue arrow. The boundaries of the three regions can be controlled varying $X_2$ and $d$. 

In the following, the emphasis is placed on most practical cases, that is, dispersive impedance sheets. According to the resonance property of sheets, the investigated structures are divided into two categories, which are non-resonant and resonant impedance sheets, respectively. Different structures corresponding to the property of impedances in each category are listed in Table~\ref{table}. The non-resonant dispersive impedance sheets will be discussed in Section~\ref{sec:NonResonant} and the resonant dispersive impedance sheets will be studied in Section~\ref{sec:resonant}.

%\begin{table}
%\textcolor{red}{\caption{List of Investigated Waveguide Structures}}
%\label{table}
%\setlength{\tabcolsep}{15pt}
%\begin{tabular}{|c|c|}
%%{|p{100pt}|p{100pt}|}
%\hline
%Property of Resonance & Property of Impedance \\
%\hline
%\multirow{3}{*}{Non-resonant}
%& Two inductive sheets \\ &Two capacitive sheets \\ & Inductive-capacitive waveguide \\
%\hline
%\multirow{3}{*}{Resonant}
%& Series-series waveguide \\ & Parallel-parallel waveguide \\ & Series-parallel waveguide \\
%\hline
%\end{tabular}
%\label{tab:list}
%\end{table}

\begin{table}
\caption{List of Investigated Waveguide Structures}
\setlength{\tabcolsep}{8pt}
\begin{tabular}{|c|c|c|}
%%{|p{100pt}|p{100pt}|}
\hline
Type & Impedance & Number of $f_{\rm{cut-off}}$\\
\hline
\multirow{3}{*}{Non-resonant}
& Inductive-Inductive & -- \\ & Capacitive-Capacitive & One for TE\\ & Inductive-Capacitive & One for TE\\
\hline
\multirow{3}{*}{Resonant}
& Series-Series & Two for TM, \par One for TE\\ & Parallel-Parallel & Two for TE\\ & Series-Parallel& One for TM, \par Two for TE\\
\hline
\end{tabular}
\label{table}
\end{table}

%%%%%%%%%%%%%%%%%%%%%%%%%%%%%%%%%%%%%%%%%%%%%%%%%%%%%%%%%%%%%%%%%%%%%%%%%%%%%%%%%%%%%%%%%%%%%%%%%%%%%%%%%%%%%%%%%%%%%%%%%%%%%%%%%%%%%%%%%%%%
\section{Non-Resonant Dispersive Impedance Sheets}\label{sec:NonResonant}

\begin{figure}[t!]	
	\centerline{\includegraphics[width=1\columnwidth]{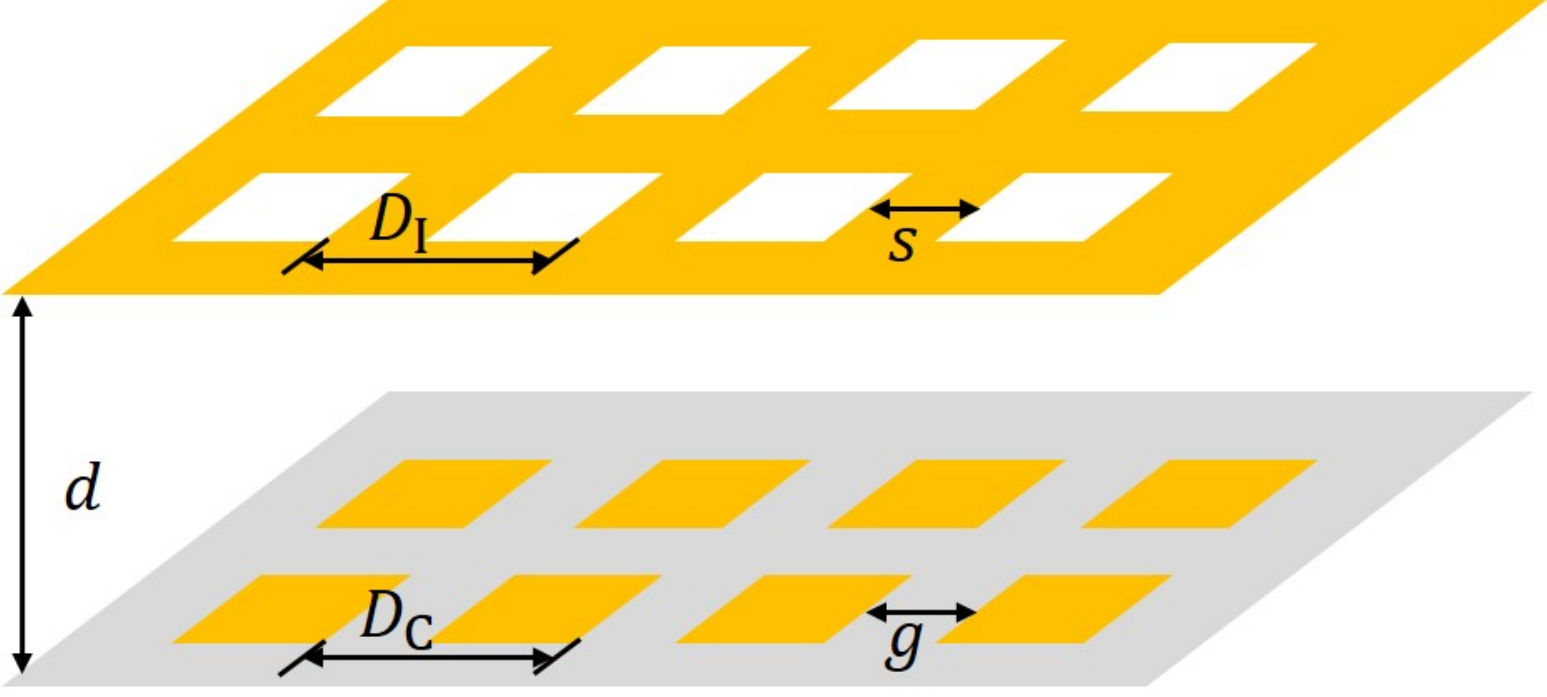}}
	\caption{The waveguide structure formed by capacitive and inductive grids.}
	\label{fig:FSSIC}
\end{figure}

Previously, we have studied the propagating modes assuming some fixed values of the sheet impedances. Here we look at the dispersion curves for metasurfaces with realistic frequency dispersion of surface reactances. First we consider capacitive and inductive sheets, that is, assume that the sheet reactances are far from resonances. Actual implementations correspond, for example, to dense meshes of metal strips for inductive sheets and complementary arrays of small metal patches  for capacitive metasurfaces. This configuration is illustrated in Fig.~\ref{fig:FSSIC}. 
%Based on the theory developed above, there can be three specific scenarios. Regarding the first and second scenario, both sheets can be grid or patch-array. The third scenario is the parallel-metasurface waveguide consisted of one grid and one patch-array layers, which is named as GP waveguide for abbreviation. In the following, we derive the dispersion relation and investigate the different scenarios that can happen. In addition, similar to previous section, the two extreme scenarios are studied.

Analytical expressions for the sheet impedances of inductive grid and patch-array are well known~\cite{FSS1,FSS2}:  
\begin{equation}
\begin{split}
&Z_{\rm{I}}=-j{\omega\mu_0 D_{\rm{I}}\over 2\pi}\ln\Big[{\displaystyle\sin({\pi s\over 2D_{\rm{I}}})}\Big]\Big(1-{\sin^2\theta\over 2}\Big),
\cr
&Z_{\rm{C}}=j{\pi\over 2\omega\epsilon_0 D_{\rm{C}}\ln\Big(\displaystyle\sin({\pi g\over 2D_{\rm{C}}})\Big)},
\label{eq:tmz1}
\end{split}
\end{equation}
for TM polarization and 
\begin{equation}
\begin{split}
&Z_{\rm{I}}=-j{\omega\mu_0 D_{\rm{I}}\over 2\pi}\ln\Big(\displaystyle\sin({\pi s\over 2D_{\rm{I}}})\Big),\cr
&Z_{\rm{C}}=j{\pi\over 2\omega\epsilon_0 D_{\rm{C}} \ln\Big(\displaystyle\sin({\pi g\over 2D_{\rm{C}}})\Big)\Big(1-{\sin^2\theta\over2}\Big)},
\label{eq:z1}
\end{split}
\end{equation}
for TE polarization.  
Here, $D_{\rm{I}}$ and $D_{\rm{C}}$ are the periods of the strip grid and the patch array,  respectively, $s$ is the strip width, and $g$ is the gap between two patches, as shown in Fig.~\ref{fig:FSSIC}. Furthermore, $\theta$ is the incidence angle. Although these surfaces are spatially dispersive (the impedances depend on the incidence angle $\theta$, that is, on the tangential component on the wavevector), we have found that for studying guided modes in considered waveguides we can substitute  $\theta=90^\circ$ maintaining high accuracy.  

From \eqref{eq:tmz1}
we see that for the considered case of electrically dense meshes (the period is small compared to the wavelength) the  sheet impedances are either inductive or capacitive.  
For compactness, it is convenient to denote $Z_{\rm{I}}=j\omega L$ and $Z_{\rm{C}}=-j/\omega C$, where 
$L$ and $C$ can be calculated using \eqref{eq:tmz1}. 
It is worth  noting that since we consider periodical metasurfaces with a subwavelength period, therefore, no radiation losses appear~\cite{FSS1}. Dissipative losses in metasurfaces can be taken into account by adding small real parts to reactive (purely imaginary) sheet impedances modeling the homogenized properties of metasurfaces. Except vicinity of resonances, losses due to finite conductivity of metals (in the microwave range) are quite small. Certainly, the scenario will be different in the near infrared region and at the visible range because we cannot avoid the dissipative losses. It is possible to use Drude model for metals and see the effect of such plasmonic losses on the propagation constant at those frequencies. However, in this paper  we consider the microwave regime in which the dissipative losses are indeed negligible.

\subsection{Two inductive sheets}

\begin{figure}[t!] 	
	\centerline{\includegraphics[width=0.6\columnwidth]{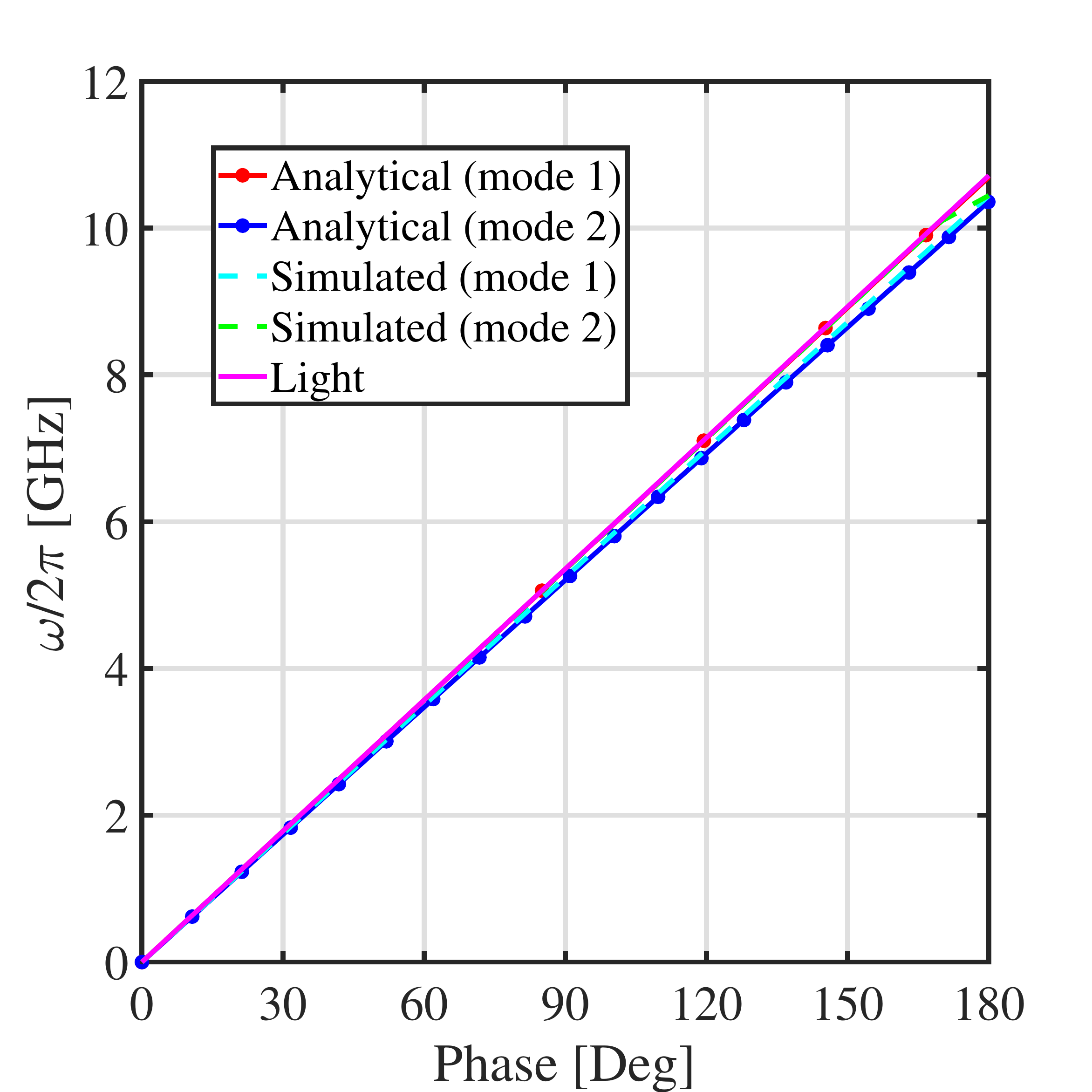}}
	\caption{Dispersion curves for the waveguide formed by two inductive sheets. Here, $D_{\rm{I}}=7$~mm, $s=3$~mm, and $d=10$~mm.}
	\label{fig:2gridFSS}
\end{figure}

When a parallel-metasurface waveguide is formed by two inductive sheets, the waveguide supports only  TM-polarized waves. Denoting by $L_2$ and $ L_3$ the surface inductances of the sheets, we write the  impedances of two sheets as $Z_2=j\omega L_2$ and $Z_3=j\omega L_3$.  Substituting $Z_2$ and $Z_3$ in \eqref{eq:gedisrelz3z2}, the dispersion relation becomes
\begin{equation}
\alpha^2\Big(e^{-2\alpha d}-1\Big)=-2\epsilon_0\omega^2\Big(L_2+L_3\Big)\alpha+4\epsilon_0^2\omega^4L_2L_3.
\label{eq:2grid}
\end{equation}
%for TM polarization.
Figure~\ref{fig:2gridFSS} presents a comparison of analytical results and numerical simulations for the structure formed by two identical inductive metasurfaces. The simulation is done by CST Microwave Studio. The results are in perfect agreement.

%%%%%%%%%%%%%%%%%%%%%%%%%%%%%%%%%%%%%%%%%%%%%%%%%%%%%%%%%%%%%%%%%%%%%
\subsection{Two capacitive sheets}

Similarly to two inductive layers, the dispersion relations of the parallel-metasurface waveguide formed by two capacitive layers can be obtained writing $Z_2=1/j\omega C_2$ and $Z_3=1/j\omega C_3$.  Substituting $Z_2$ and $Z_3$ in \eqref{eq:DisR}, the dispersion relation (TE modes) reads
\begin{equation}
{4\alpha^2\over C_2 C_3}-2 \mu_0 \alpha \Big({1\over C_2}+{1\over C_3}\Big)\omega^2+\mu_0^2 (1-e^{-2\alpha d})\omega^4=0.
\label{eq:2capactive}
\end{equation}
% for TE polarization. 
 One of the modes  has a cut-off at 
\begin{equation}
f_{\rm{cut-off}}\approx{{1\over 2\pi}{\sqrt{C_2+C_3\over \mu_0 d C_2 C_3}}}.
\label{eq:cutoff capacitance}
\end{equation}
%%%%%%%%%%%%%%%%%%%%%%%%%%%%%%%%%%%%%%%%%%%%%%%%%%%%%%%%%%%%%%%%%%%%%

\subsection{Inductive-capacitive waveguide}

When a parallel-metasurface waveguide consists of one inductive grid and one capacitive grid, replacing $Z_2=j\omega L$ and $Z_3=1/(j\omega C)$ in \eqref{eq:gedisrelz3z2} and \eqref{eq:DisR} gives the following dispersion relations:
\begin{equation}
{2\epsilon_0\alpha\over C}+\alpha^2(1-e^{-2\alpha d})=\omega^2\Big[{4\epsilon_0^2 L\over C}+2\epsilon_0\alpha L\Big],
\label{eq:TMFSS}
\end{equation}
for the TM polarization and 
\begin{equation}
{4\alpha^2 L\over C}+{2\mu_0\alpha\over C}=\omega^2\Big[2\mu_0\alpha L+\mu_0^2(1-e^{-2\alpha d})\Big],
\label{eq:TEFSS}
\end{equation}
for the TE polarization. For this polarization, there is a cut-off frequency which is estimated as
\begin{equation}
f_{\rm{cut-off}}\approx{{1\over 2\pi}{1\over\sqrt{C(L+\mu_0 d)}}}.
\label{eq:TEFSS cutoff}
\end{equation}

\begin{figure}[ht!]	
	\centerline{\includegraphics[width=0.6\columnwidth]{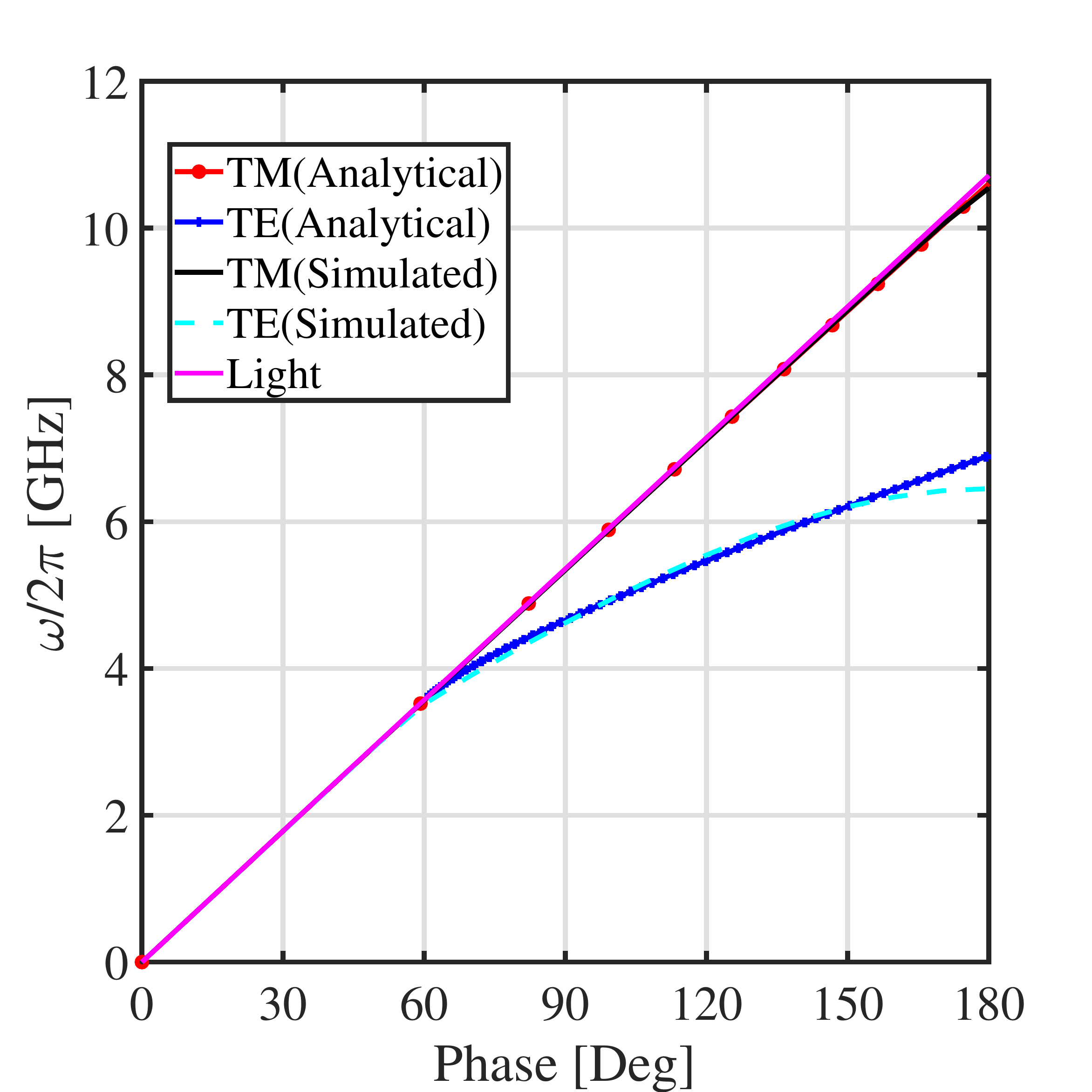}}
	\caption{Dispersion curves for the waveguide formed by one capacitive and one inductive sheet. Here, $D_{\rm{I}}=7$~mm, $D_{\rm{C}}=14$~mm, $s=3$~mm, $g=0.2$~mm, and $d=10$~mm.}
	\label{fig:FSS}
\end{figure}

In order to examine the accuracy of the results, analytical results and numerical simulations are compared in Fig.~\ref{fig:FSS}. The results are in perfect agreement. The dispersion curve of the TM mode is close to the light line, while the curve for the TE mode is far from the light line. The cut-off frequency of the TE mode is approximately $3.57$~GHz.

\begin{figure*}[t!]	\centering
	\subfigure[]{\includegraphics[width=0.25\textwidth]{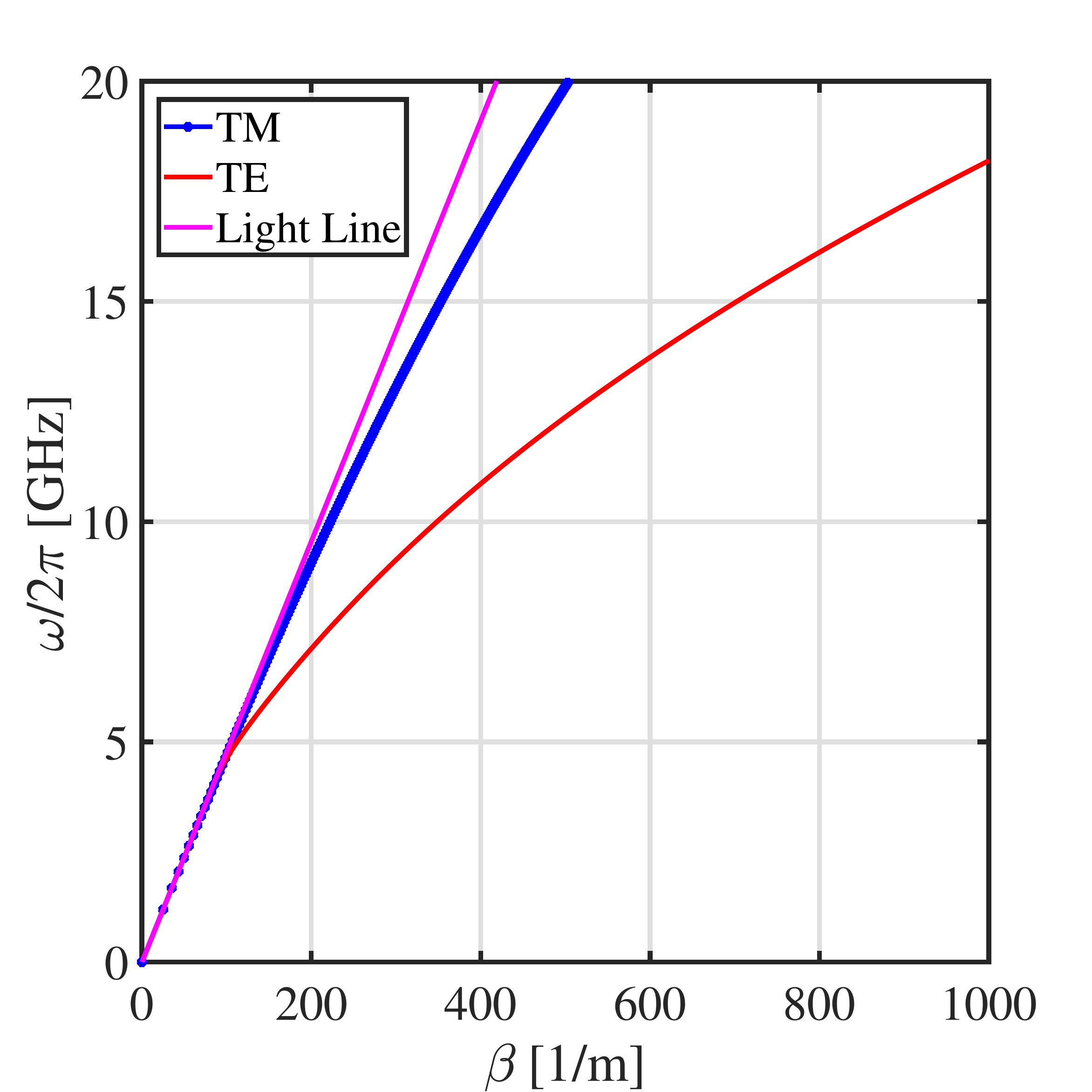}}
	\subfigure[]{\includegraphics[width=0.25\textwidth]{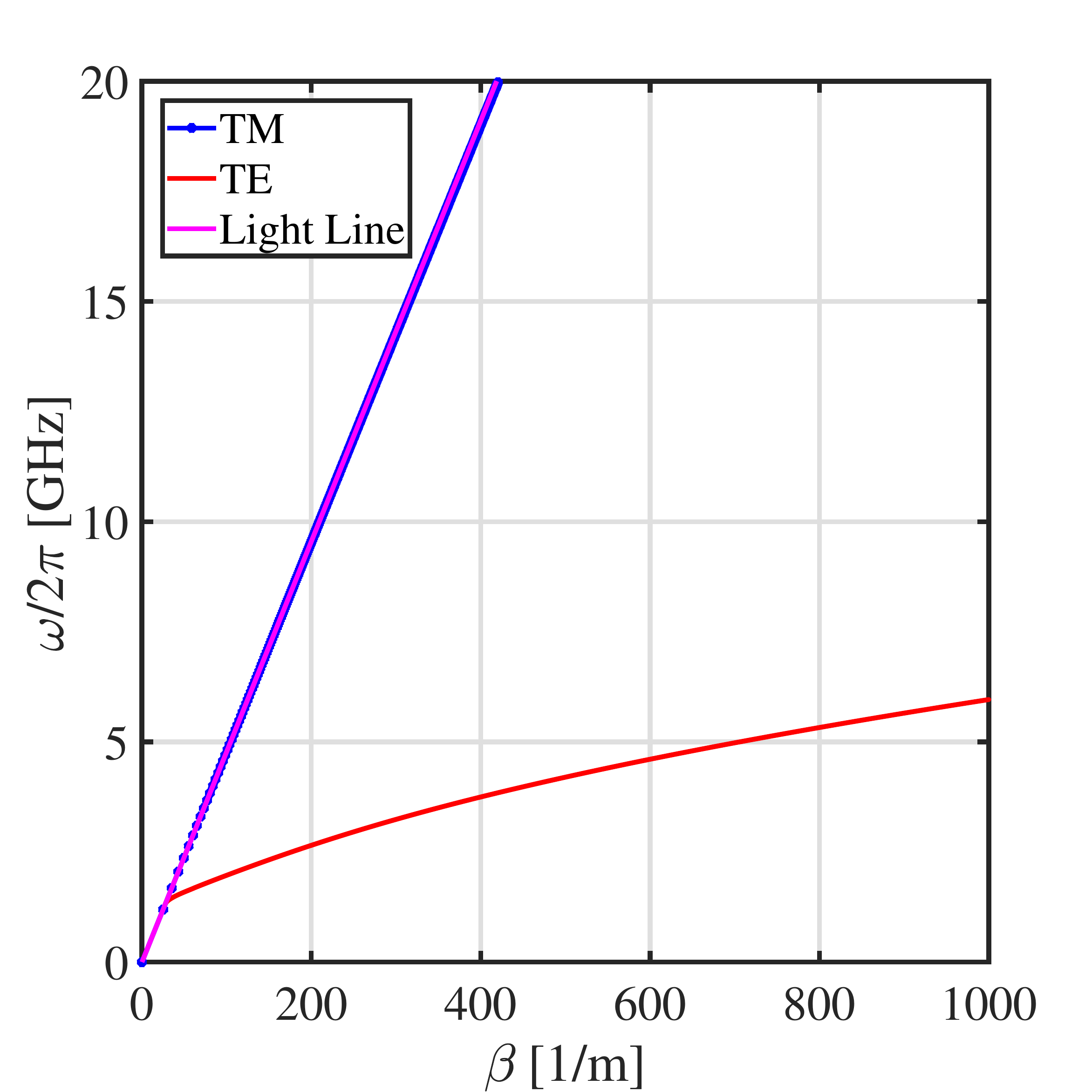}}
	\subfigure[]{\includegraphics[width=0.25\textwidth]{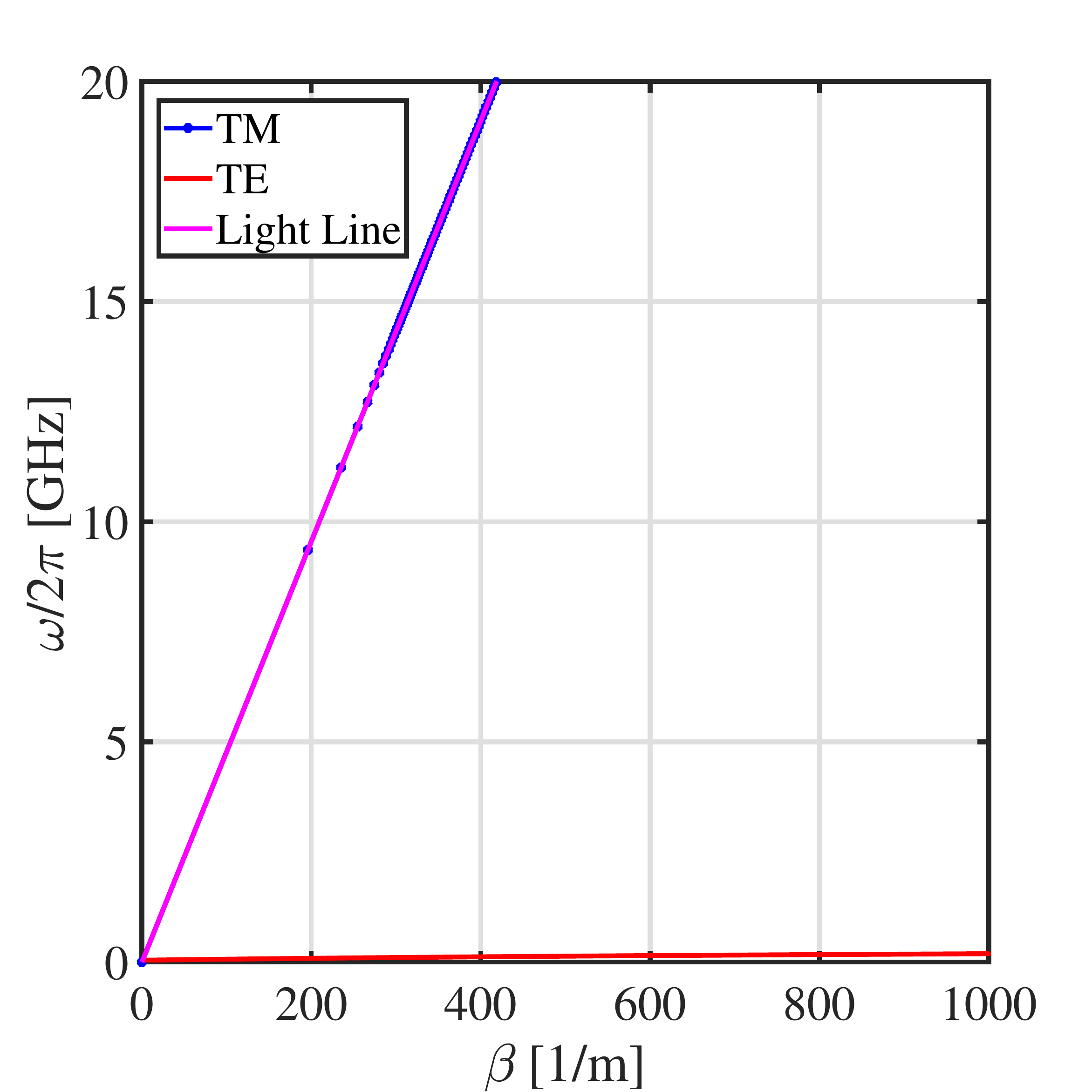}}
	\caption{Dispersion curves of the structure in~\cite{F} which keeps $Z_2=-Z_3$ at the frequency where the two reactances form a parallel resonance. Here, $f_0=15$~GHz, $d=10$~mm, (a) $Z_2=j94$~$\Omega$, $L=1$~nH, and $C=0.1126$~pF. (b) $Z_2=j9.4$~$\Omega$, $L=0.1$~nH, and $C=1.126$~pF. (c) $Z_2=j0.0094$~$\Omega$, $L=0.0001$~nH, and $C=1126$~pF.}
	\label{fig:SamePhaseFSS}
\end{figure*}
Interestingly, after careful investigations of extreme scenarios for non-resonant dispersive sheets, we find the same distribution of dispersion curve
with Fig.~\ref{fig:ExteTM} and~\ref{fig:ExteTE}.

Recently, parallel arrays of reactive sheets have been proposed for realization of exotic reflection/transmission properties. Such structures to some extent resemble realizations of bound states, and of interest for invisibility and cloaking applications~\cite{F,Krasnok}.
In \cite{F}, two parallel effectively homogeneous reactive sheets were considered. At the frequency where the two reactances form a parallel resonant structure and the separation between the two sheets is $m\lambda/2$ ($m=1,2,\dots$), the fields outside the volume are not perturbed by the metasurfaces while the fields between the sheets can be controlled. In addition, there is an efficient route to excite bound states in the continuous spectrum of modes with theoretically infinite lifetime is to dynamically tune the reactance of both metasurfaces to extremely low values keeping the balance $Z_2=-Z_3$~\cite{F}. Here, we analyze the eigenmodes of this structure to understand what guided modes can be excited.

Figure~\ref{fig:SamePhaseFSS} shows the dispersion curves of the structure in~\cite{F} for dramatically different values of the reactance of the two metasurfaces. We keep the distance between two sheets $\lambda/2$. At $f_0=15$~GHz, where the transmission coefficient is unity for normal incidence, it is simultaneously possible to excite TE and TM modes at the input port of the waveguide. For both modes, the energy is confined inside the structure and attenuates outside in free space. However, the attenuation constant is quite different for different modes. The TE mode has a higher attenuation constant compared to the TM mode. Notice that the TE mode always has a cut-off frequency approximately given by \eqref{eq:TEFSS cutoff}, while the TM mode does not have any cut-off. Interestingly, in the limit of infinitely small inductance of one sheet and infinitely large capacitance of the other, the dispersion curve of the TM mode approaches the light line, while the dispersion curve of TE mode collapses to zero frequency. In analogy with our parallel-plate waveguide operating at this limit, there is an open structure (so-called planar ENZ-dielectric-ENZ waveguide) introduced in Ref.~\cite{Monticone}. That  structure is a dielectric slab covered by two plasmonic low-loss layers. Under TM-polarized plane-wave illumination, at the plasma frequency in which the permittivity of the plasmonic layers is near zero, and finally under the condition that the plasma frequency coincides with the Fabry-Perot resonance of the dielectric slab, it was shown that the quality factor of this resonant open structure diverges resulting in a  bound mode~\cite{Monticone}. This phenomenon is similar with what we found in Ref.~\cite{F}.

%%%%%%%%%%%%%%%%%%%%%%%%%%%%%%%%%%%%%%%%%%%%%%%%%%%%%%%%%%%%%%%%%%%%%%%%%%%%%%%%%%%%%%%%%%%%%%%%%%%%%%%%%%%%%%%%%%%%%%%%%%%%%%%%%%%%%%%%%%%
\section{Resonant Dispersive Impedance}
\label{sec:resonant}
\begin{figure}[ht!]
	\subfigure[]{\includegraphics[width=0.33\columnwidth]{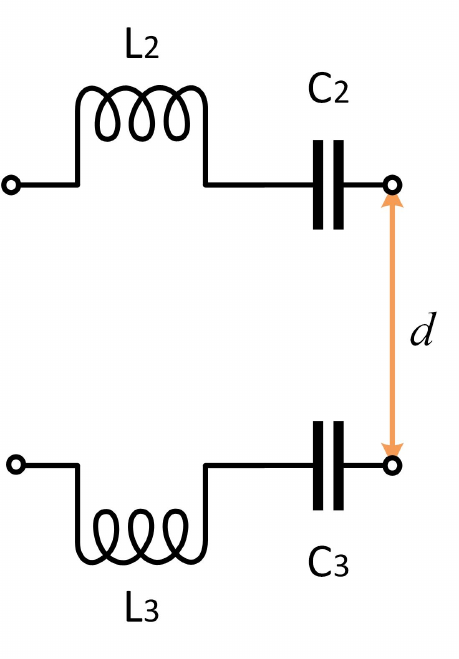}}\subfigure[]{\includegraphics[width=0.33\columnwidth]{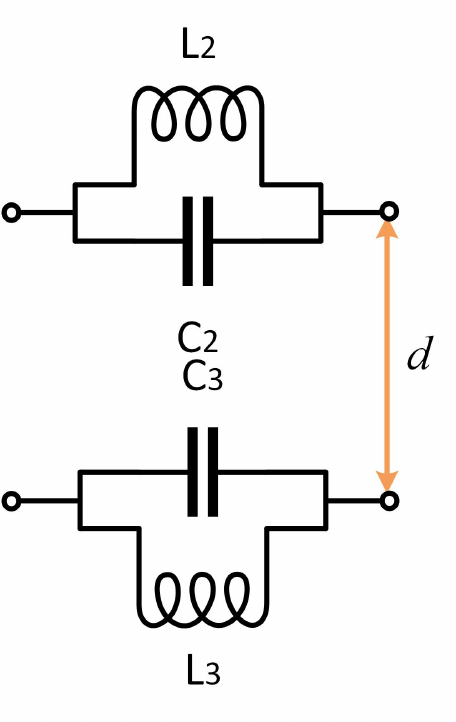}}
	\subfigure[]{\includegraphics[width=0.33\columnwidth]{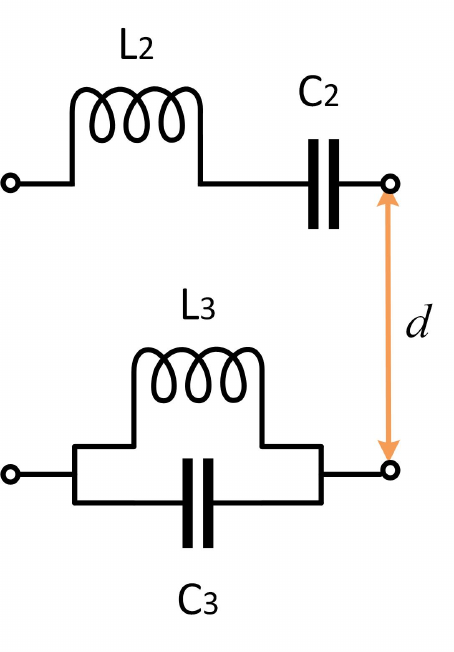}}
	\caption{Three specific scenarios for resonant impedance waveguides in which (a) the impedances of both sheets are series-resonant, (b)the impedances of both sheets are parallel-resonant, and (c) the impedance of one sheet is a series and the other one is a parallel circuit.}
	\label{fig:circuit}
\end{figure}
\begin{figure*}[t!]\centering
	\subfigure[]{\includegraphics[width=0.25\textwidth]{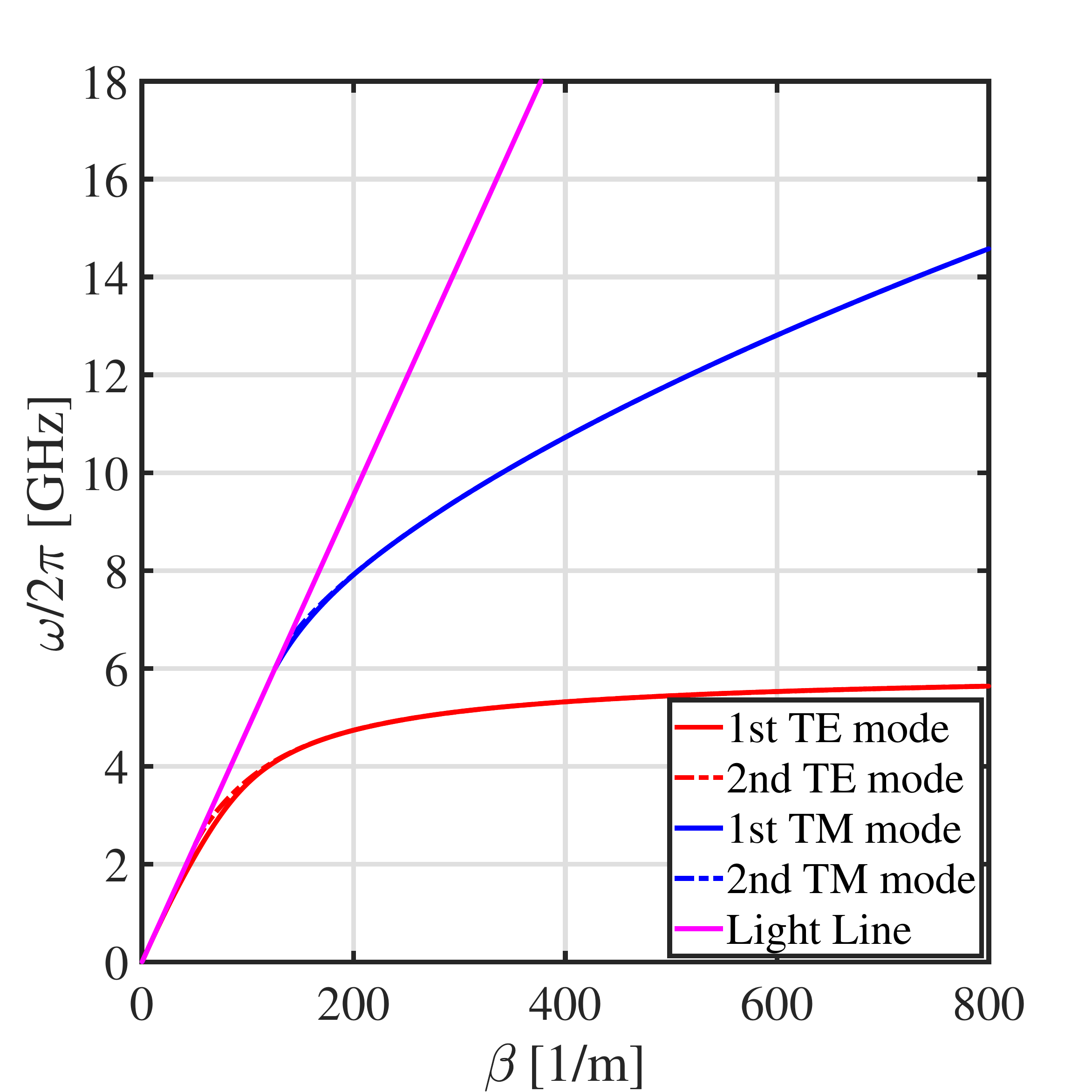}}
	\subfigure[]{\includegraphics[width=0.25\textwidth]{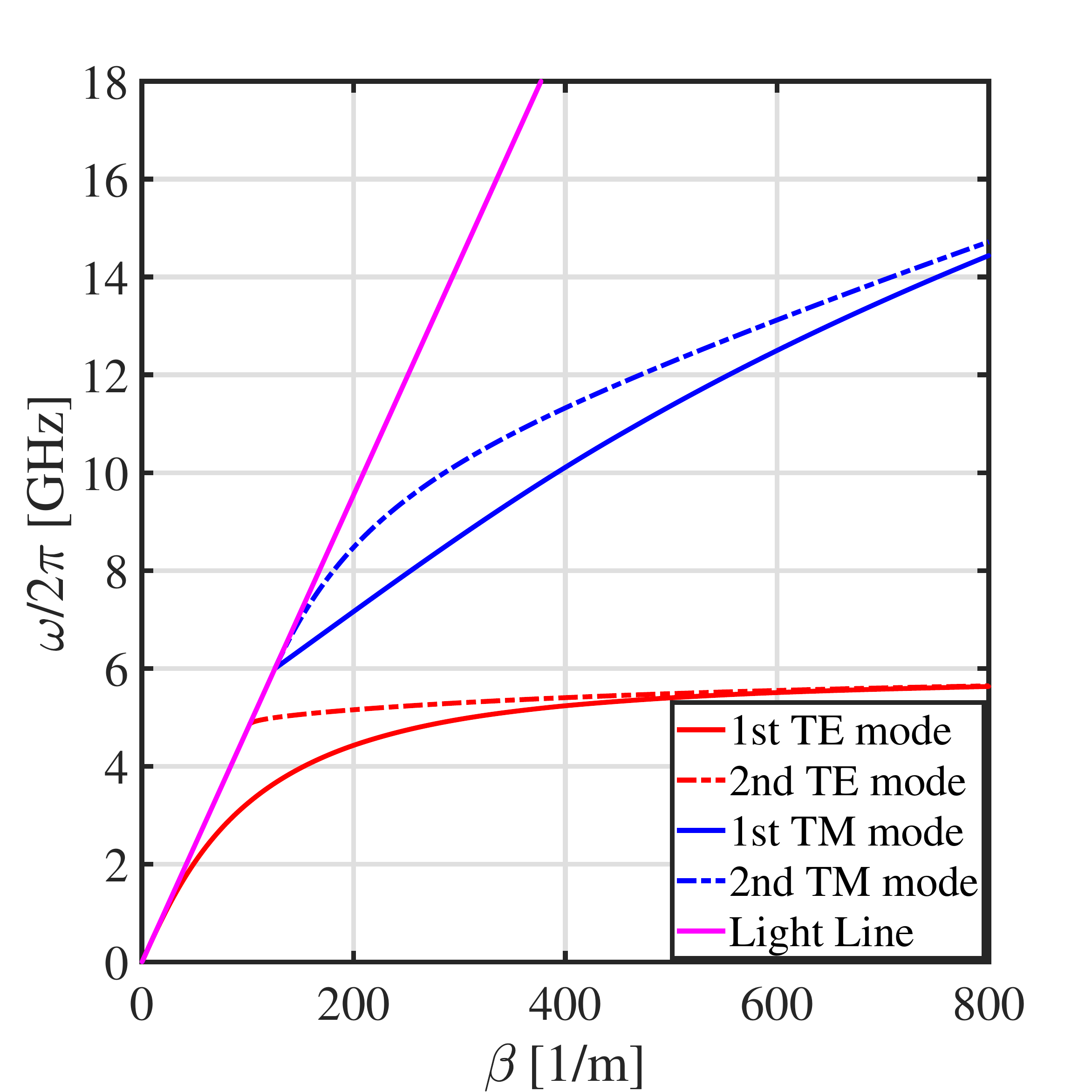}}
	\subfigure[]{\includegraphics[width=0.25\textwidth]{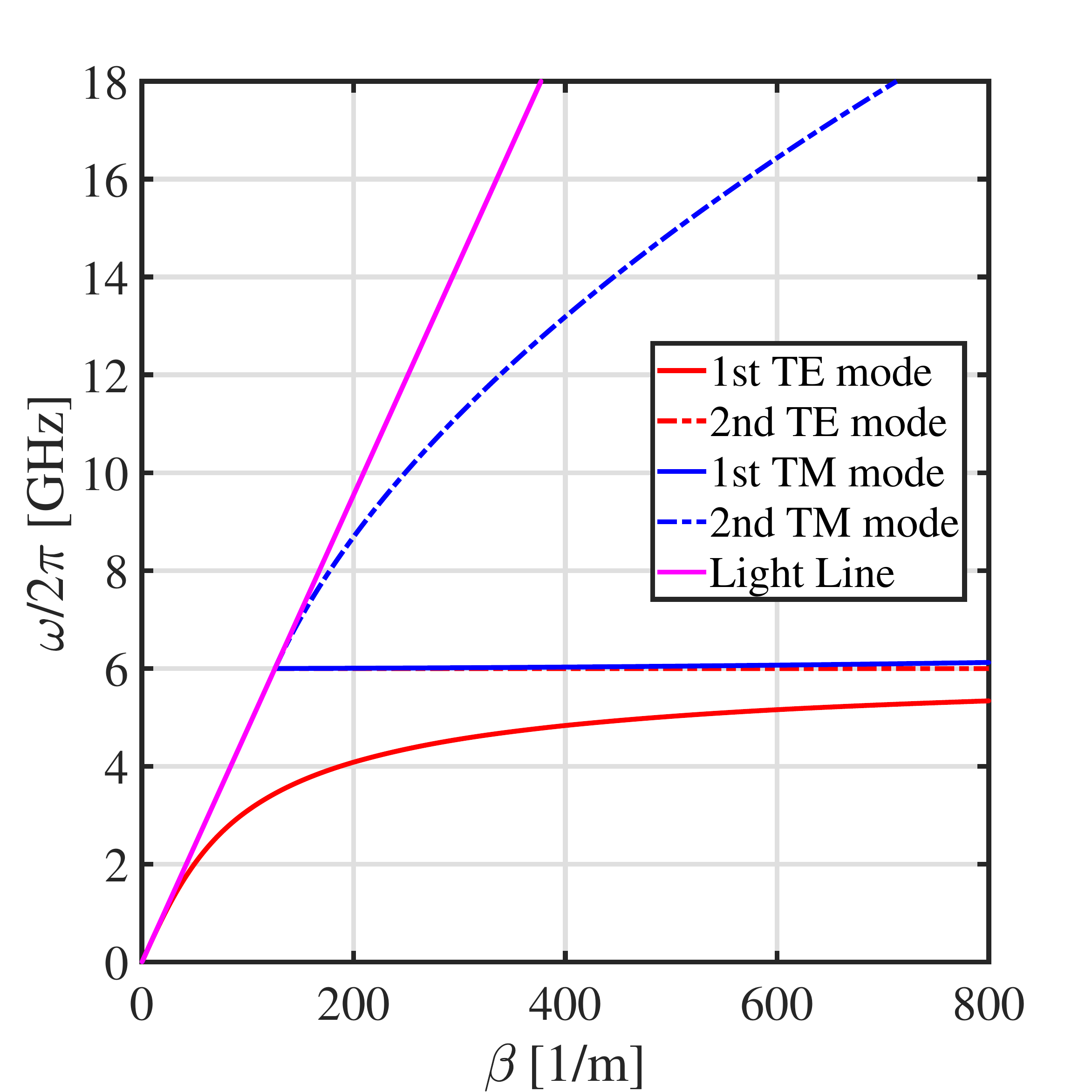}}	
	\caption{Dispersion curves for a symmetric structure for series connections in which the distances are (a) $d=\lambda_{\rm{6GHz}}$, (b) $d=\lambda_{\rm{6GHz}}/10$, and (c) $d=\lambda_{\rm{6GHz}}/5000$.}
	\label{fig:similar series}
\end{figure*} 
\begin{figure*}[h!]\centering
	\subfigure[]{\includegraphics[width=0.25\textwidth]{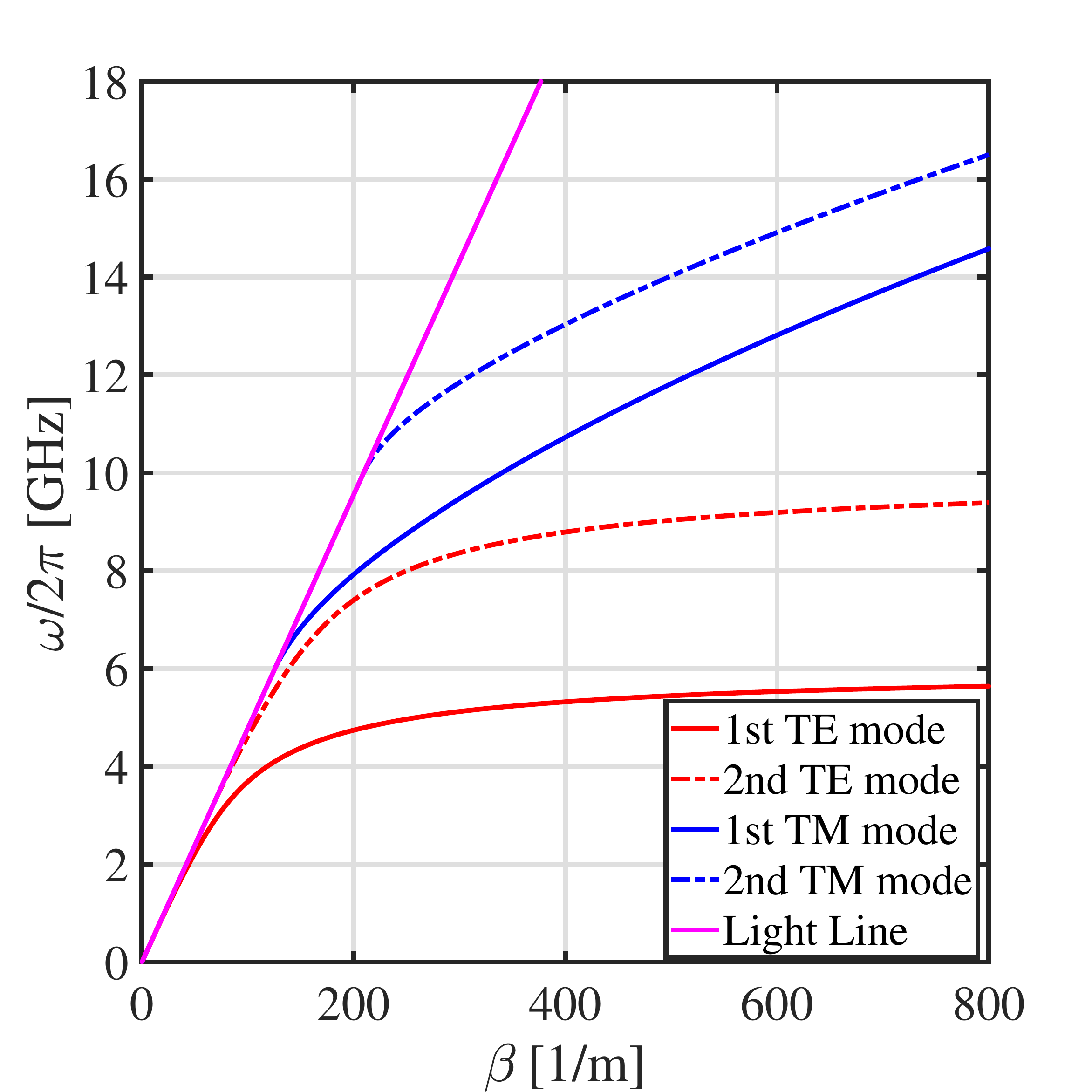}}
	\subfigure[]{\includegraphics[width=0.25\textwidth]{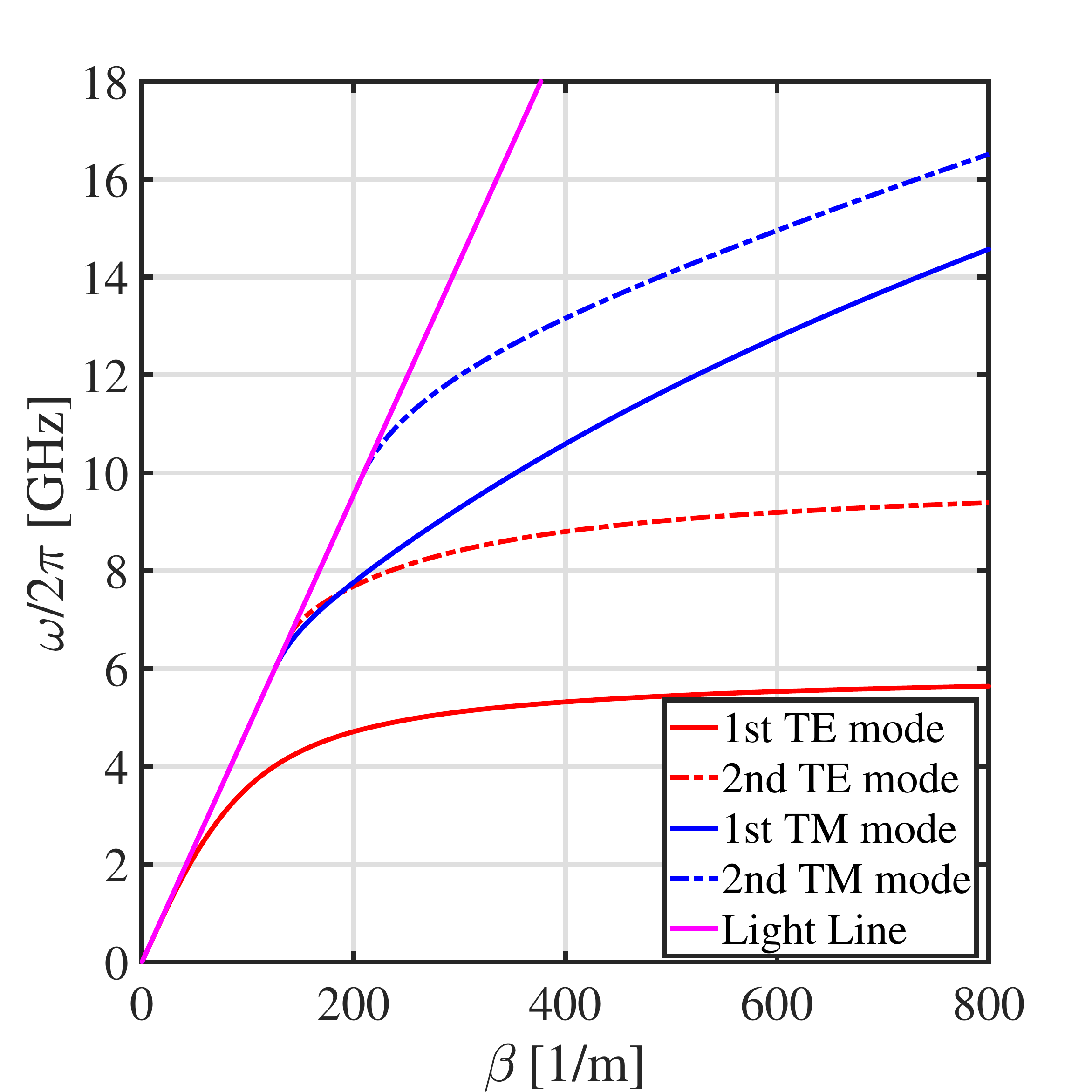}}
	\subfigure[]{\includegraphics[width=0.25\textwidth]{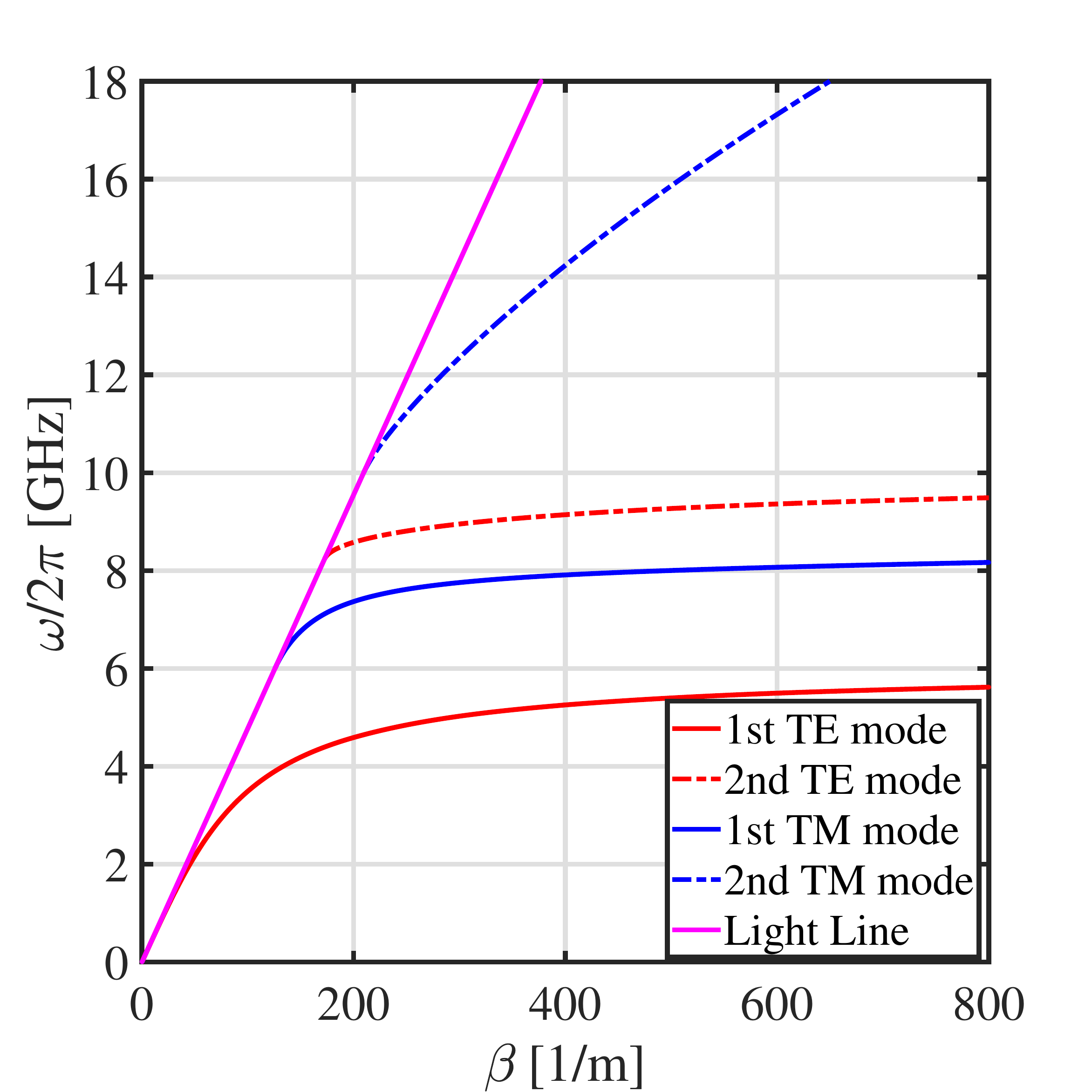}}
	\caption{Dispersion curves for an asymmetric structure for series connections in which the distances are (a) $d=\lambda_{\rm{6GHz}}$, (b) $d=\lambda_{\rm{6GHz}}/10$, and (c) $d=\lambda_{\rm{6GHz}}/5000$.}
	\label{fig:different series}
\end{figure*} 
In recent work~\cite{Krasnok}, the bound mode in a double array of identical small lossless resonant particles was identified. Here, we study propagation along two effectively resonant sheets in which the impedance can be the realized by a series connection or a parallel connection of sheet reactances and provide physical insight into the behavior of surface waves when the distance between the sheets is very small. Since we have two sheets, there can be three specific scenarios as illustrated by Fig.~\ref{fig:circuit}. Concerning the first and second scenarios, the impedances of both sheets exhibit series or parallel resonances. The third scenario is when the impedance of one sheet has a series resonance while the other one has a parallel resonance. The homogenized sheet impedance is characterized by effective inductance and capacitance (denoted as $L_2$, $C_2$ and $L_3$, $C_3$ for each sheet) which are obtained from the full-wave solution of a plane-wave reflection problem in the quasi-static limit and can be calculated from the formulas in~\cite{self-resonant,self-resonant2} for self-resonant structures. The resonance frequencies of two sheets are determined by $f_{2}=1/(2\pi\sqrt{C_2 L_2})$ and $f_{3}=1/(2\pi\sqrt{C_3 L_3})$, respectively. For each scenario, two different cases need to be considered: Asymmetric and symmetric. In the symmetric case the sheets are the same ($Z_2=Z_3$), while in the asymmetric case, the two sheets possess different impedances ($Z_2\neq Z_3$).

%%%%%%%%%%%%%%%%%%%%%%%%%%%%%%%%%%%%%%%%%%%%%%%%%%%%%%%%%%%%%%%%%%%%%
\subsection{First scenario: Sheet impedances as series connections of reactances}
\label{subsec:series}

The waveguide consists of two sheets in which the impedances exhibit series resonances. We write the sheet impedances as  $Z_2=j\omega L_2+1/j \omega C_2$ and $Z_3=j\omega L_3+1/j \omega C_3$.  Substituting $Z_2$ and $Z_3$ in \eqref{eq:gedisrelz3z2} and \eqref{eq:DisR}, the TM-mode dispersion relation for the first scenario can be written as
\begin{equation}
\begin{split}
4\epsilon_0^2 L_2 L_3\omega^4-\Big[4\epsilon_0^2({L_2\over C_3}+{L_3\over C_2})+2 \alpha \epsilon_0 (L_2+L_3) \Big]\omega^2+\cr
2\epsilon_0\alpha({1\over C_3}+{1\over C_2})+{4\epsilon_0^2\over C_2 C_3}+ \alpha^2\Big(1-e^{-2\alpha d}\Big)=0.
\end{split}
\end{equation}
As mentioned above, two TM modes can survive within the structure and both of them have cut-off frequencies which we denote as $f_{\rm{cut-off}}=f_{2}$ and $f_{\rm{cut-off}}=f_{3}$. These modes are dramatically different from the non-resonant scenario (discussed in Section~\ref{sec:NonResonant}) which have no cut-off frequency for two inductive sheets. The dispersion relations for TE polarization reads:
\begin{equation}
\begin{split}
\Big[4\alpha^2L_2L_3+2\mu_0\alpha (L_2+L_3)+\mu_0^2\Big(1-e^{-2\alpha d}\Big)\Big]\omega^4-\cr
\Big[4\alpha^2 ({L_2\over C_3}+{L_3\over C_2})+2\mu_0\alpha ({1\over C_3}+{1\over C_2})\Big]\omega^2+{4 \alpha^2 \over C_2 C_3}=0.
\end{split}
\end{equation}
For this polarization, two proper solutions can be obtained. Notably, one of the TE modes suffers from a cut-off frequency which relates to the distance between the two sheets and can be calculated by:
\begin{equation}
f_{\rm{cut-off}}\approx{{1 \over 2 \pi} \sqrt{C_2+C_3\over C_2 C_3 (L_2+L_3+\mu_0 d)} }.
\label{eq:fcut series}		
\end{equation}

As an example, $6$~GHz is chosen as the resonance frequency of both impedances ($f_2=f_3=6$~GHz), in which $L_2=6$~nH, $C_2=0.11727$~pF. For two different impedance sheets, the resonance frequency of one sheet is set to $10$~GHz ($f_2=6$~GHz, $f_3=10$~GHz). In this case, we assume that $L_2=6$~nH, $C_2=0.11727$~pF, $L_3=6$~nH, and $C_3=0.042217$~pF. The analytical results for different distances are shown in Fig.~\ref{fig:similar series} and  Fig.~\ref{fig:different series}.

Regarding the case of symmetric sheets separated by $d=\lambda_{\rm{6GHz}}$, two TE modes exist below the resonance frequency, while two TM modes can propagate above the resonance frequency, since here both sheets are inductive. The two TE (or TM) modes have approximately the same phase velocity in the symmetric case. Interestingly, decreasing the distance results in separation of two TE (or TM) modes in which one of them ``rotates'' in the direction closer to the resonance frequency. The limit of $d\rightarrow 0$ leads to the existence of only one TE and one TM mode. Note that if the distance between the sheets becomes comparable or smaller than the period of the structures forming the sheets, the homogenized models of the two metasurfaces in terms of their sheet impedances become not enough adequate due to significant coupling between the sheets via evanescent fields of structural inhomogeneities. More advances analytical models or  full-wave simulations are needed in this case.

\begin{figure}[t!]\centering
	\subfigure[]{\includegraphics[width=0.15\textwidth]{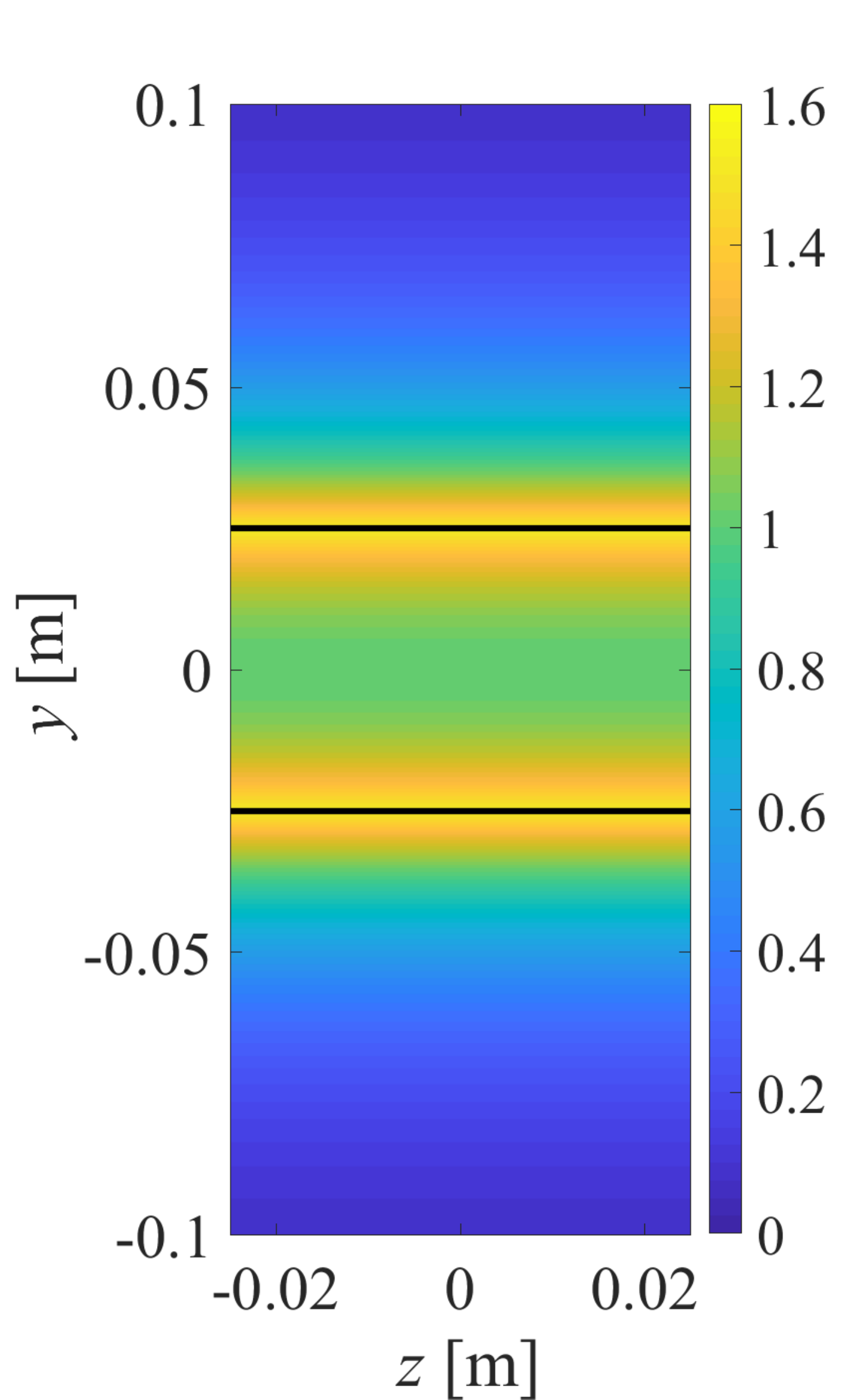}}
	\subfigure[]{\includegraphics[width=0.15\textwidth]{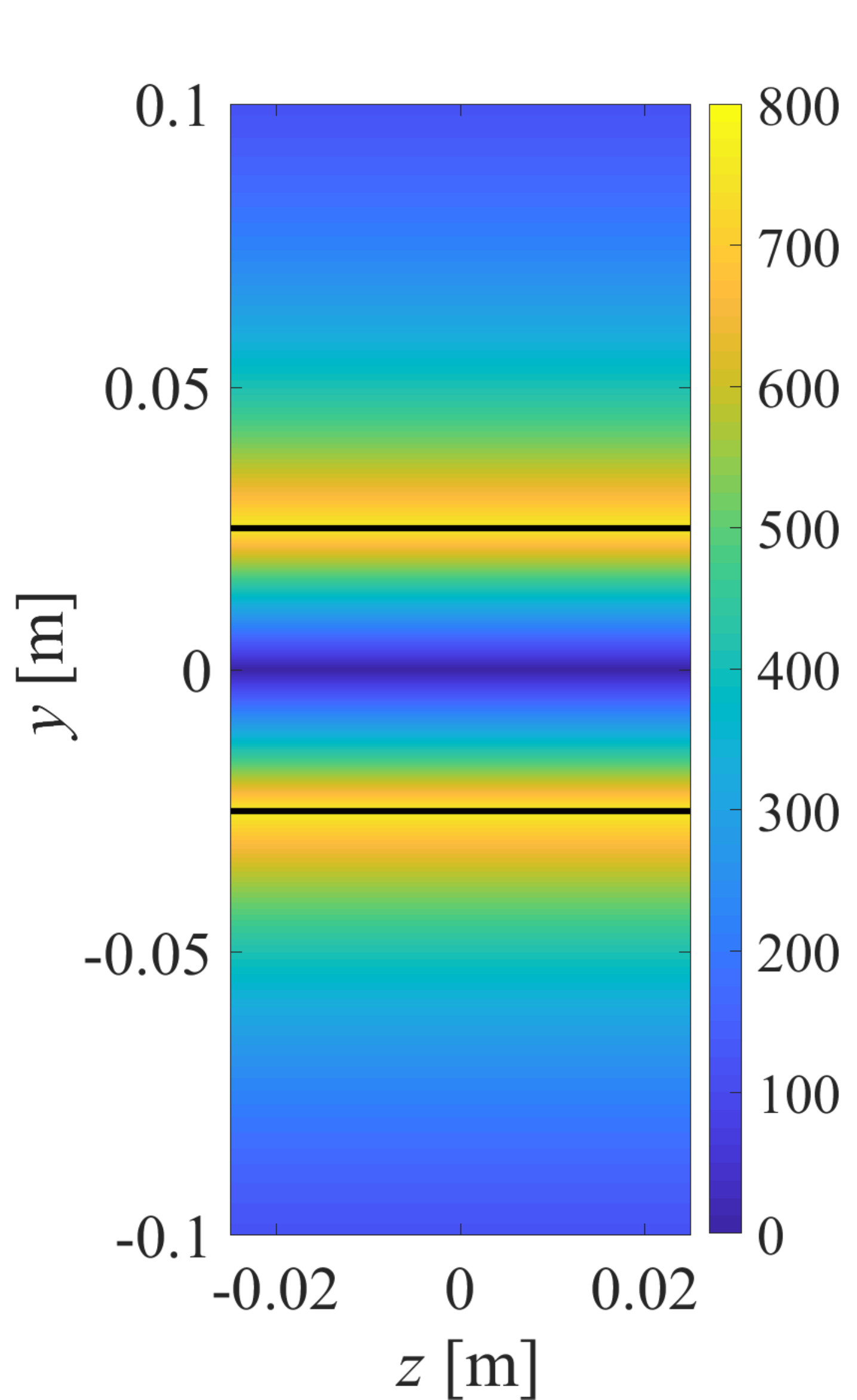}}
	\subfigure[]{\includegraphics[width=0.15\textwidth]{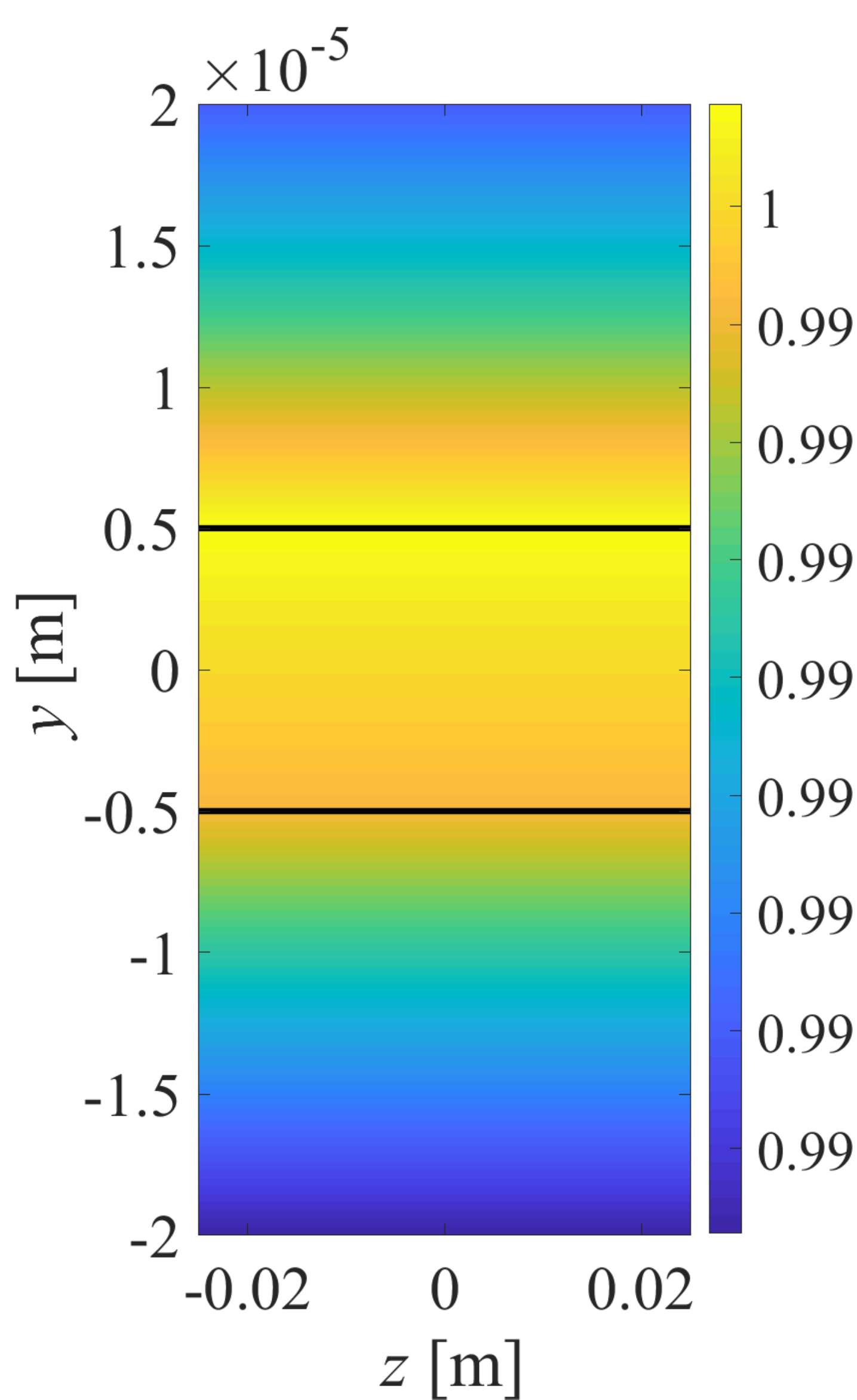}}
	\caption{Electric field distribution of symmetric structure in series for TE modes in which (a) 1st mode, $d=\lambda_{\rm{6GHz}}$, $\beta=74.3$ 1/m, (b) 2nd mode, $d=\lambda_{\rm{6GHz}}$, $\beta=67.4$ 1/m, and (c) 1st mode, $d=\lambda_{\rm{6GHz}}/5000$, $\beta=81.7$ 1/m. Here, the operational frequency is $f=3$ GHz. The vertical and horizontal axes are $y$- and $z$-axes, respectively. The metasurface positions are shown by solid black lines.}
	\label{fig:Series TE fields}
\end{figure}

\begin{figure*}[ht!]\centering
	\subfigure[]{\includegraphics[width=0.25\textwidth]{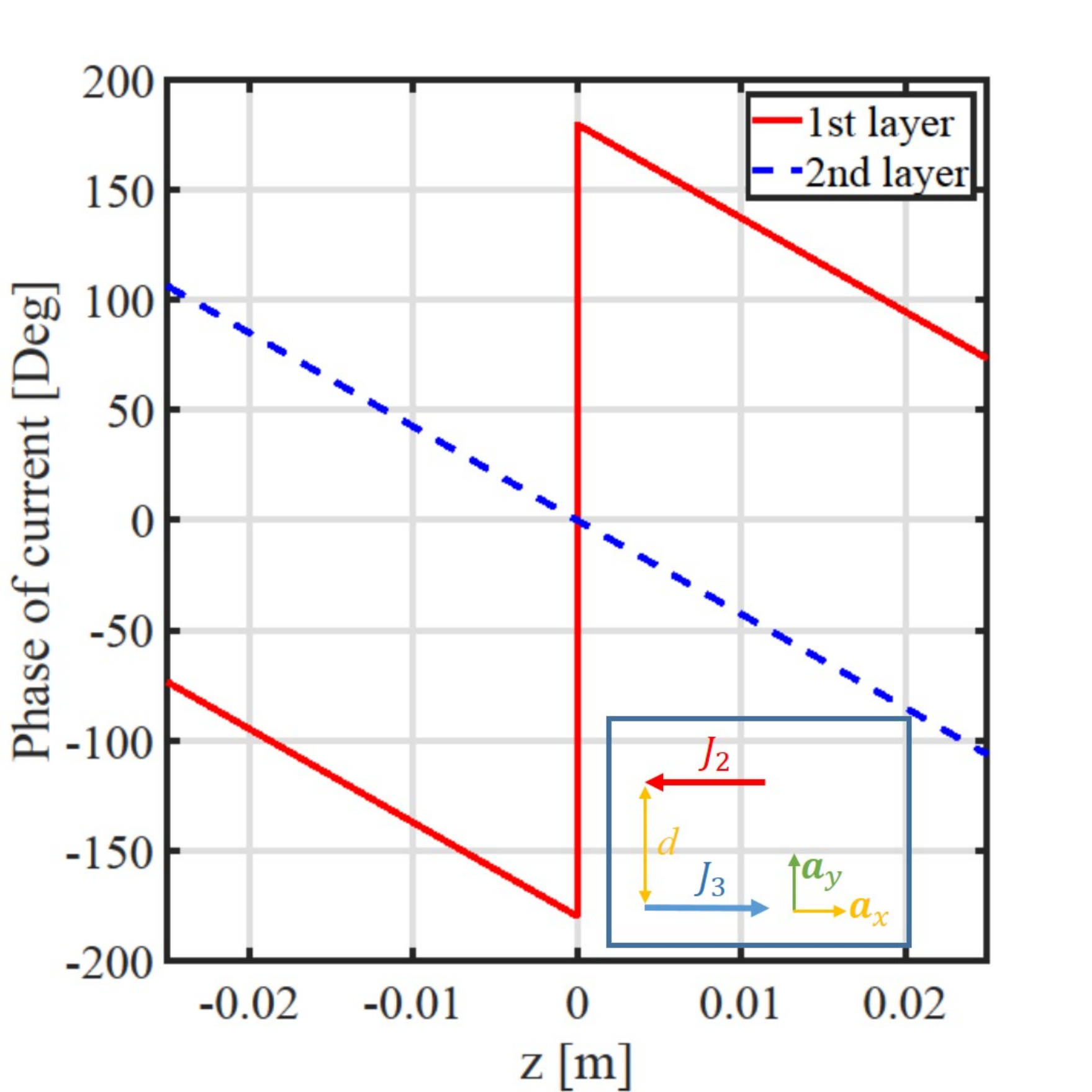}}
	\subfigure[]{\includegraphics[width=0.25\textwidth]{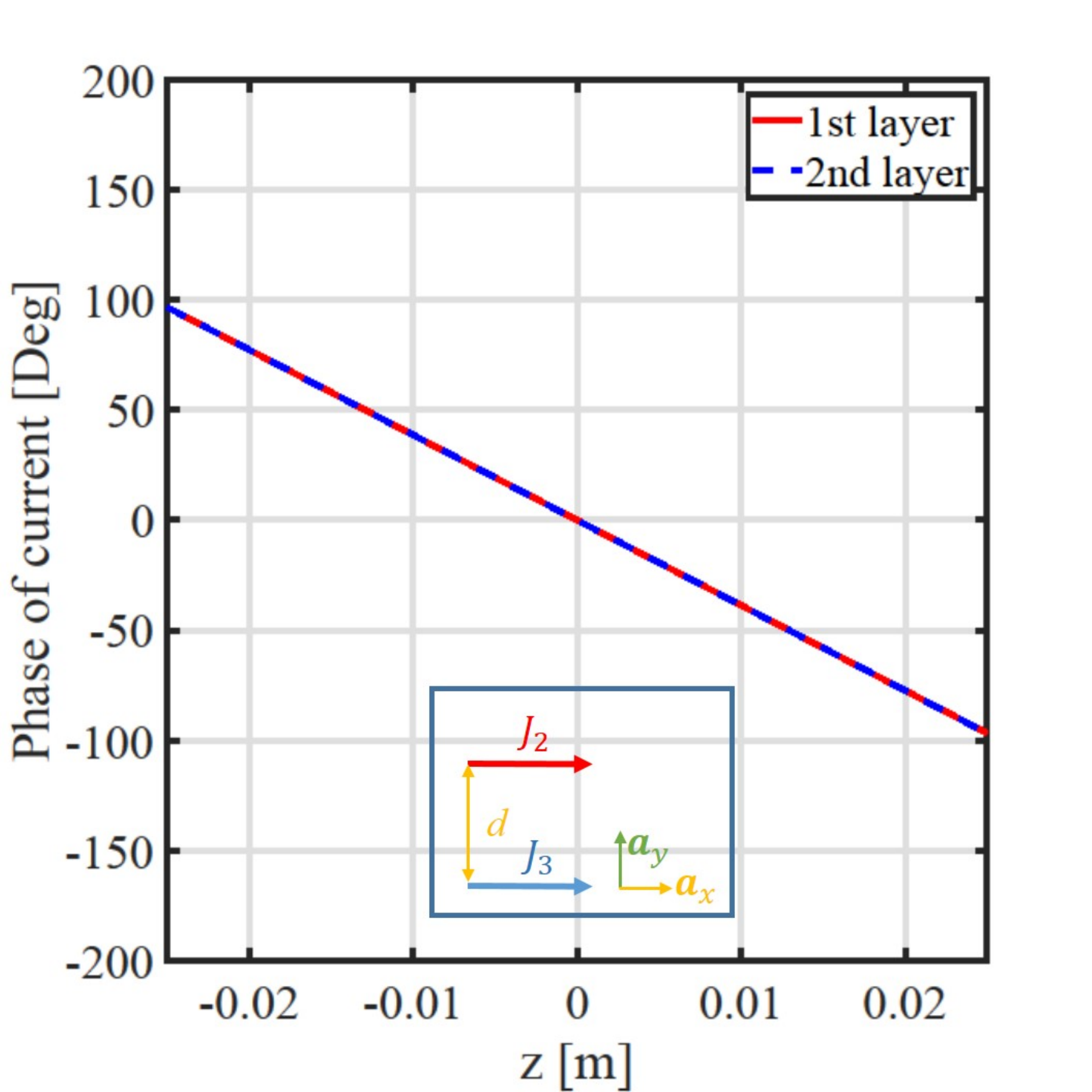}}
	\subfigure[]{\includegraphics[width=0.25\textwidth]{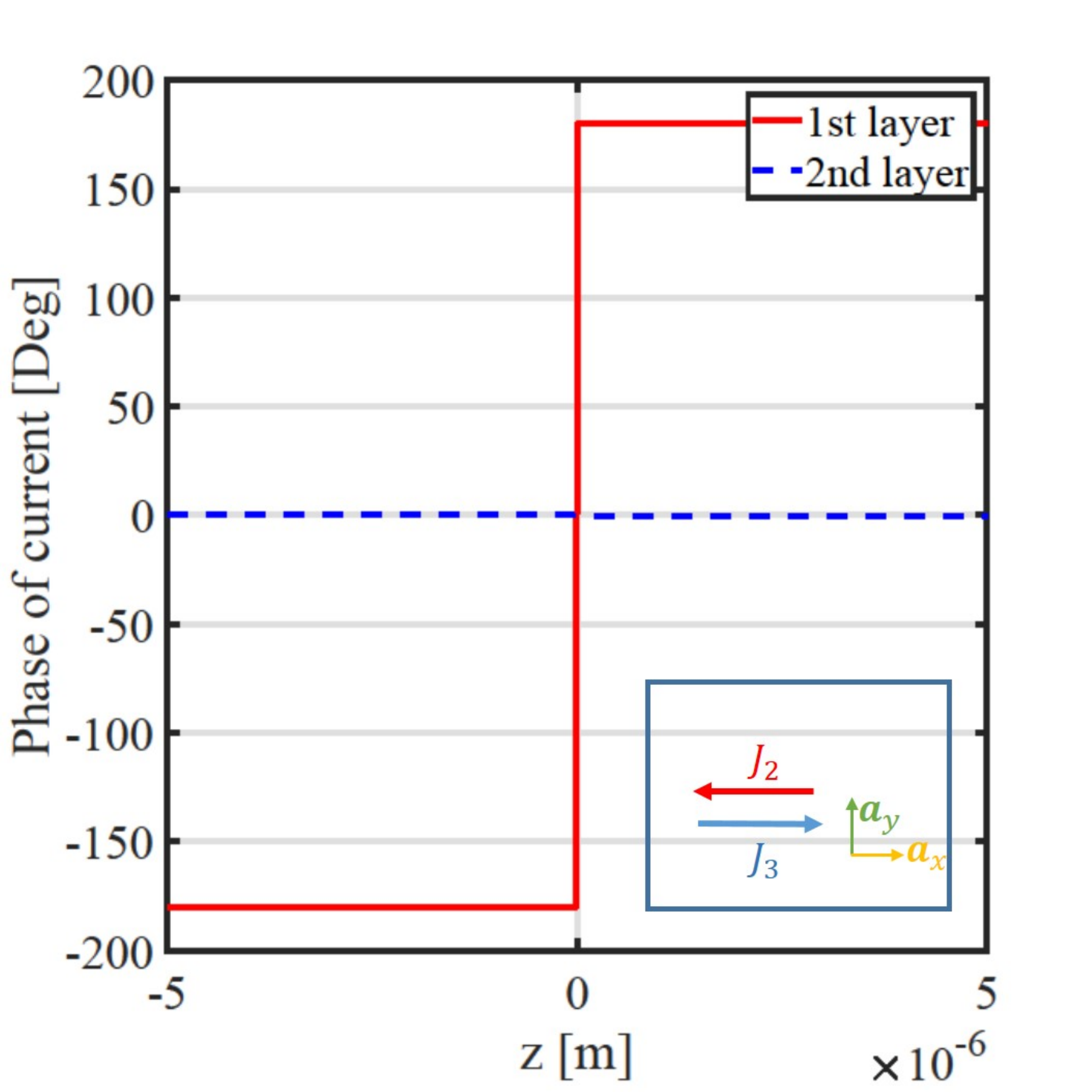}}
	\caption{Phase of surface currents in a symmetric structure in series connection for TE modes in which (a) 1st mode, $d=\lambda_{\rm{6GHz}}$, $\beta=74.3$ 1/m, (b) 2nd mode, $d=\lambda_{\rm{6GHz}}$, $\beta=67.5$ 1/m, and (c) 1st mode, $d=\lambda_{\rm{6GHz}}/5000$, $\beta=81.7$ 1/m. Here, the operational frequency is $f=3$ GHz.}
	\label{fig:Series TE current}
\end{figure*}

In order to get better physical insight into the behavior of surface modes, Fig.~\ref{fig:Series TE fields} plots the electric field distributions, and Fig.~\ref{fig:Series TE current} shows the surface current density for a symmetric structure for TE modes at $3$~GHz when $d=\lambda_{\rm{6GHz}}$ and $d=\lambda_{\rm{6GHz}}/5000$, respectively.
There are two TE modes since both sheets are capacitive at $3$~GHz. The two black solid lines show the positions of the two metasurfaces. One can see that for $d=\lambda_{\rm{6GHz}}$ the field is strongly tied to both sheets and attenuates as the vertical distance from that sheet increases. When the two sheets get close to each other, strong coupling can be observed from the phenomenon that the field does not attenuate in the vertical direction between the two sheets because of the small distance. The surface current of TE-polarized modes flows along the $x$-axis. It is perceived that the equal magnitude of surface current density on two sheets can be conveniently attained due to the equal amplitude of electric field at both sheets which have identical impedances.
Nevertheless, the phases of the surface current are dramatically different for these two modes. As it is shown in Fig.~\ref{fig:Series TE current}, when $d=\lambda_{\rm{6GHz}}$, the first mode is out-of-phase mode and the second mode is in-phase mode. When the two sheets are close to each other, the in-phase mode disappears and only the out-of-phase mode remains. 

\begin{figure*}[h!]\centering
		\subfigure[]{\includegraphics[width=0.25\textwidth]{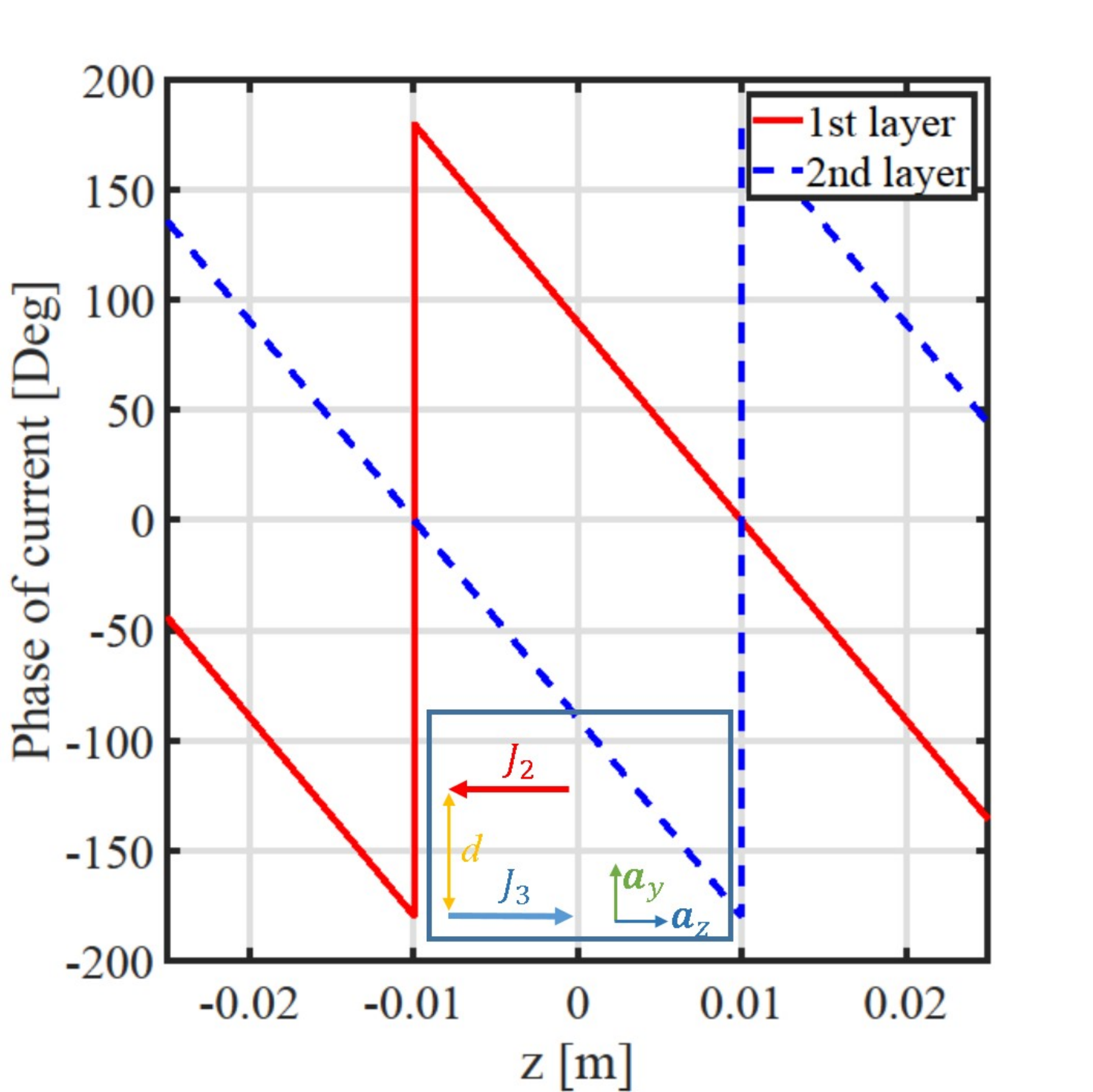}}
		\subfigure[]{\includegraphics[width=0.25\textwidth]{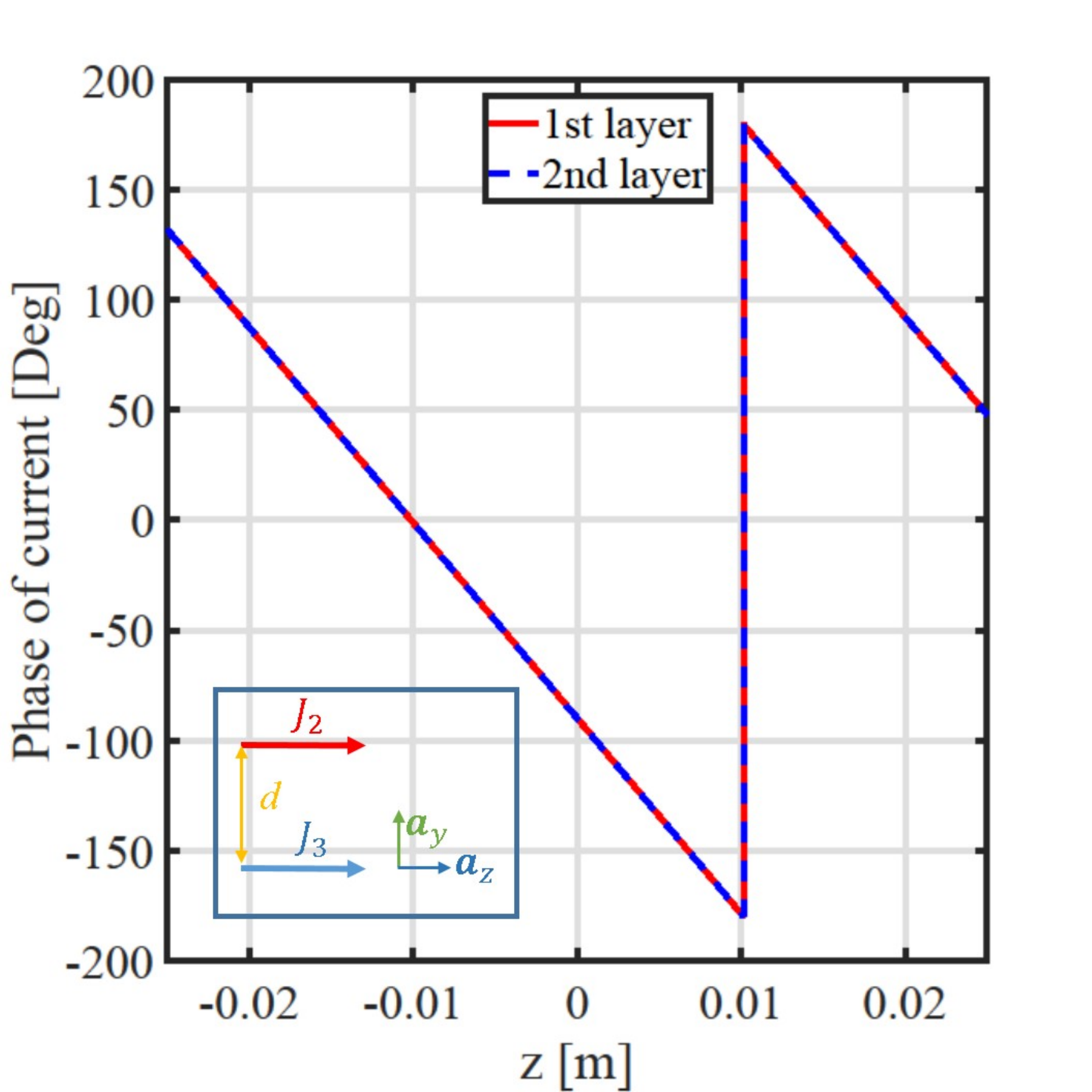}}
		\subfigure[]{\includegraphics[width=0.25\textwidth]{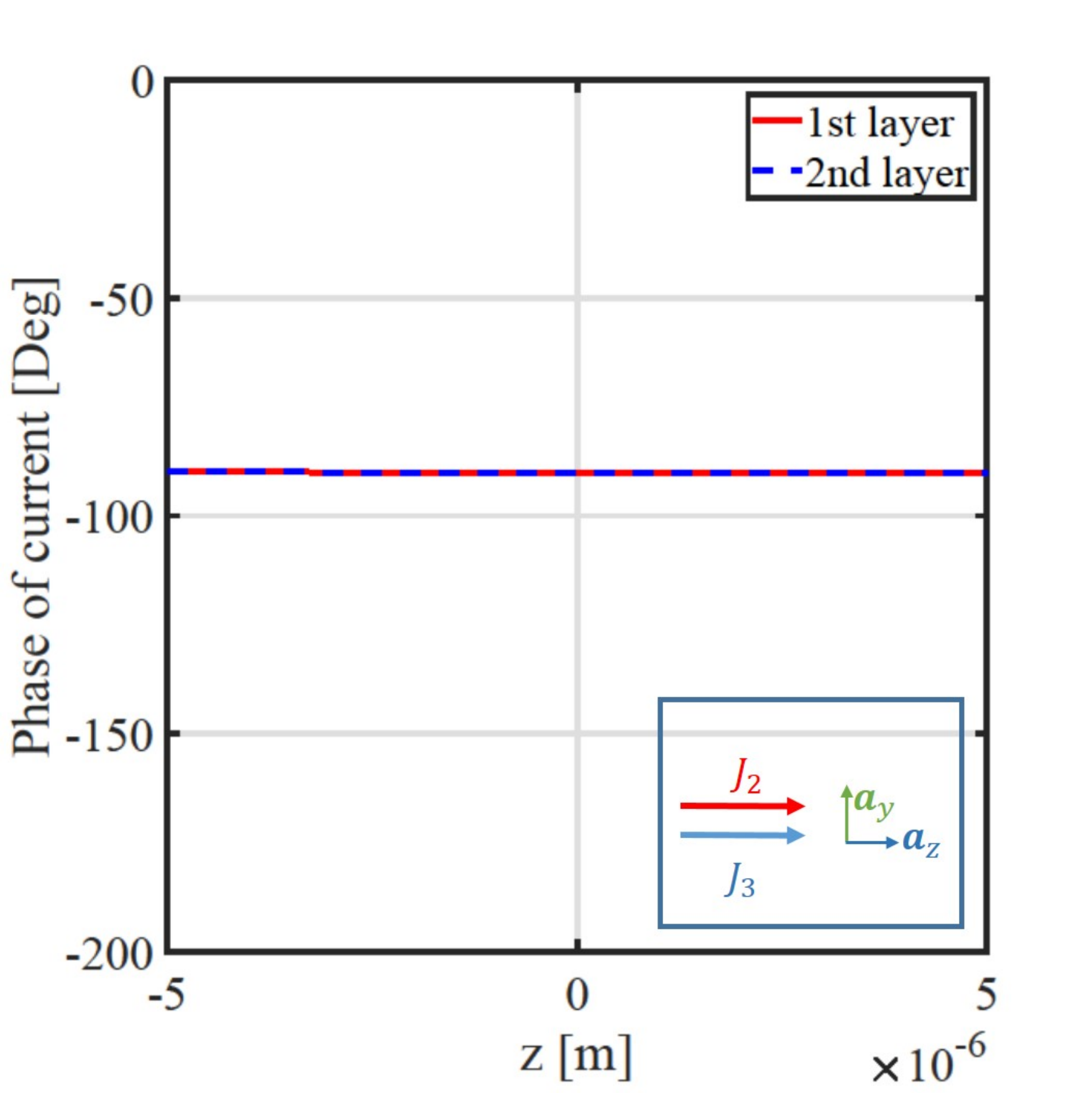}}
	\caption{Phase of surface currents in a symmetric structure in series connection for TM modes in which (a) 1st mode, $d=\lambda_{\rm{6GHz}}$, $\beta=157$ 1/m, (b) 2nd mode, $d=\lambda_{\rm{6GHz}}$, $\beta=155$ 1/m, and (c) 1st mode, $d=\lambda_{\rm{6GHz}}/5000$, $\beta=149.1$ 1/m. Here, the operational frequency is $f=7$ GHz.}
	\label{fig:Series TM current}
\end{figure*}

For the TM modes, the surface currents on the two sheets flow along the $z$-axis which is the direction of surface wave propagation. The phases of surface currents are quite different from that of the TE modes, albeit the symmetric magnitude of surface current. As it is shown in Fig.~\ref{fig:Series TM current}, in the limit of zero $d$ the double sheets can support a  TM mode, where the currents induced on the two arrays are equal and in phase. The out-of-phase mode disappears. 

Concerning the case of the asymmetric structure, as shown in Fig.~\ref{fig:different series}, two TE modes are supported below $6$~GHz since both sheets are capacitive. In the range from 6 GHz to 10 GHz, both TE and TM waves can propagate since one sheet is capacitive and the other one is inductive. Above 10 GHz, two TM modes exist because of two inductive sheets. As we bring the two sheets close to each other, the cut-off frequency of the second TE mode increases and a new resonance frequency $f_{\rm{mix}}$ emerges.  The dispersion curves are classified by $f_2$, $f_3$ and $f_{\rm{mix}}$ which is given by
\begin{equation}
f_{\rm{mix}}={{1 \over 2 \pi} \sqrt{C_2+C_3\over C_2 C_3 (L_2+L_3)} }.
\label{eq:fmix}	
\end{equation}

\begin{figure}[t!]\centerline 
{\includegraphics[width=0.6\columnwidth]{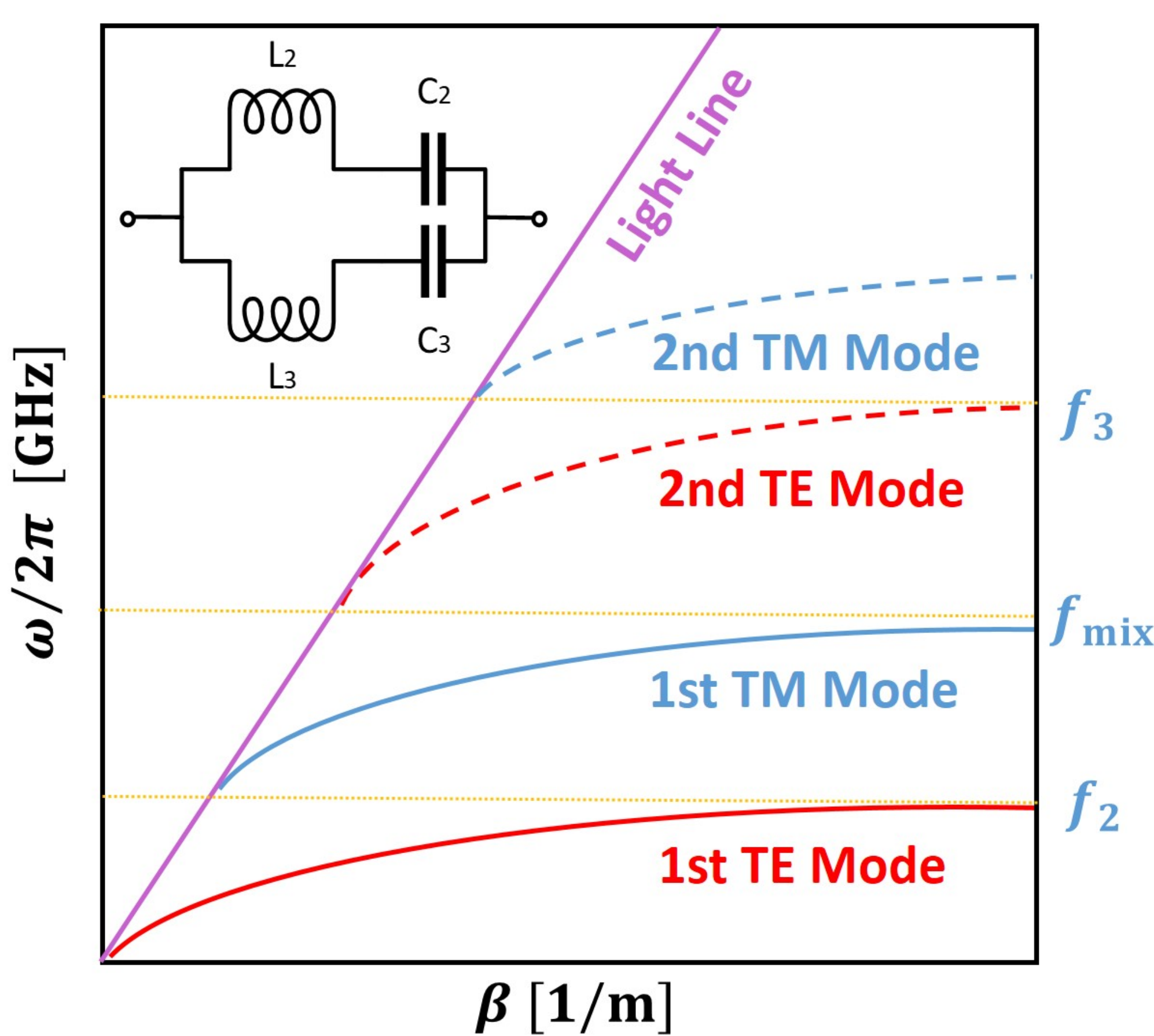}}
\caption{Dispersion curves for asymmetric sheets for series connections in which distance goes to zero.}
\label{fig:series extreme}
\end{figure}

When the distance is exactly zero, $f_{\rm{mix}}$ is the same as the cut-off frequency of the TE mode, which can be easily obtained by comparing \eqref{eq:fcut series} with \eqref{eq:fmix}. Practically, there is a very narrow stop  band at the new resonance frequency ($f_{\rm{mix}}$) for surface waves (in the region where the propagation factor for the TM wave is very large but TE waves still cannot propagate). The dispersion curves for the extreme case are illustrated in Fig.~\ref{fig:series extreme}. 

Clearly, the case of symmetric sheets corresponds to a special case of the asymmetric sheets. In the limit of zero $d$, the first TM mode and second TE mode of the symmetric structure are compressed to one resonance frequency, because $f_2=f_3=f_{\rm{mix}}$.

%%%%%%%%%%%%%%%%%%%%%%%%%%%%%%%%%%%%%%%%%%%%%%%%%%%%%%%%%%%%%%%%%%%%%
\subsection{Second
scenario: Sheet impedances as parallel connections of reactances}
\label{subsec:parallel}
The waveguide consists of two meatsufaces in which the impedances are parallel connections of reactances, expressed by $Z_2=j\omega L_2/(1-\omega^2 L_2C_2)$ and $Z_3=j\omega L_3/(1-\omega^2 L_3C_3)$. Substituting $Z_2$ and $Z_3$ in \eqref{eq:gedisrelz3z2} and \eqref{eq:DisR}, the dispersion equation for TM polarization can be found:
\begin{equation}
\begin{split}
&\Big[2\epsilon_0\alpha L_2 L_3(C_2+ C_3)+\alpha^2\Big(1-e^{-2\alpha d} \Big)L_2 L_3C_2C_3+\cr
&4\epsilon_0^2 L_2 L_3\Big]\omega^4-
\Big[2\epsilon_0\alpha(L_2+L_3)+\cr
& \alpha^2\Big(1-e^{-2\alpha d} \Big) (C_2L_2+ C_3L_3) \Big]\omega^2+\alpha^2\Big(1-e^{-2\alpha d}\Big)=0.
\end{split}
\end{equation}
For this polarization, there is no cut-off frequency. The dispersion equation for the TE polarization reads
\begin{equation}
\begin{split}
\mu_0^2\Big(1-e^{-2\alpha d}\Big)L_2 L_3C_2C_3\omega^4-
\Big[2\mu_0\alpha L_2L_3( C_2+C_3)+\cr
\mu_0^2\Big(1-e^{-2\alpha d}\Big)(C_2L_2+ C_3L_3)\Big]\omega^2+\mu_0^2\Big(1-e^{-2\alpha d}\Big)+\cr
4\alpha^2 L_2 L_3+2\mu_0\alpha (L_2+L_3)=0.
\end{split}
\end{equation}
When $Z_2=Z_3$, the cut-off frequency for this polarization is $f_{\rm{cut-off}}=f_2$ ($f_2=f_3$).

\begin{figure*}[ht!]
	\centering
	\subfigure[]{\includegraphics[width=0.24\textwidth]{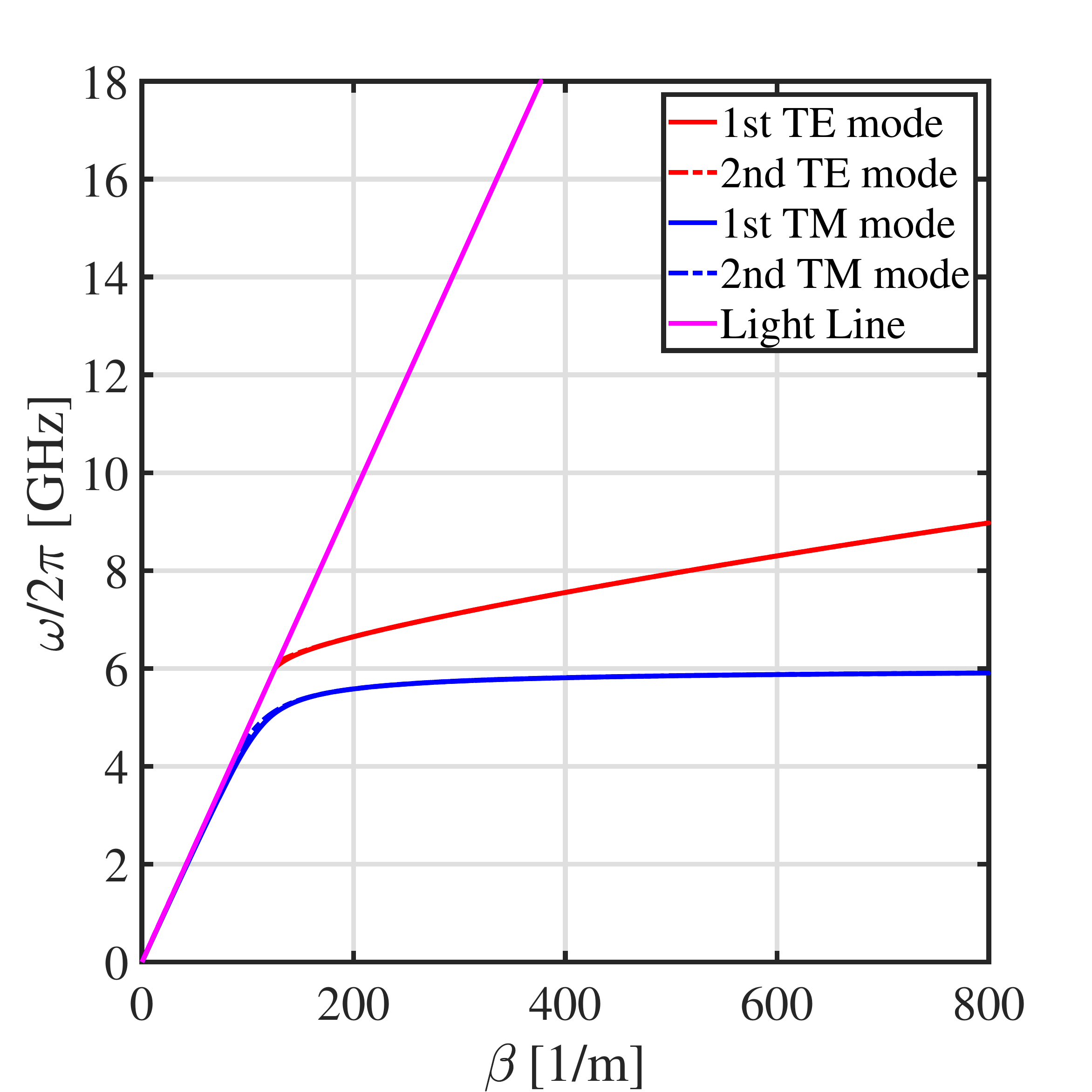}}
	\subfigure[]{\includegraphics[width=0.24\textwidth]{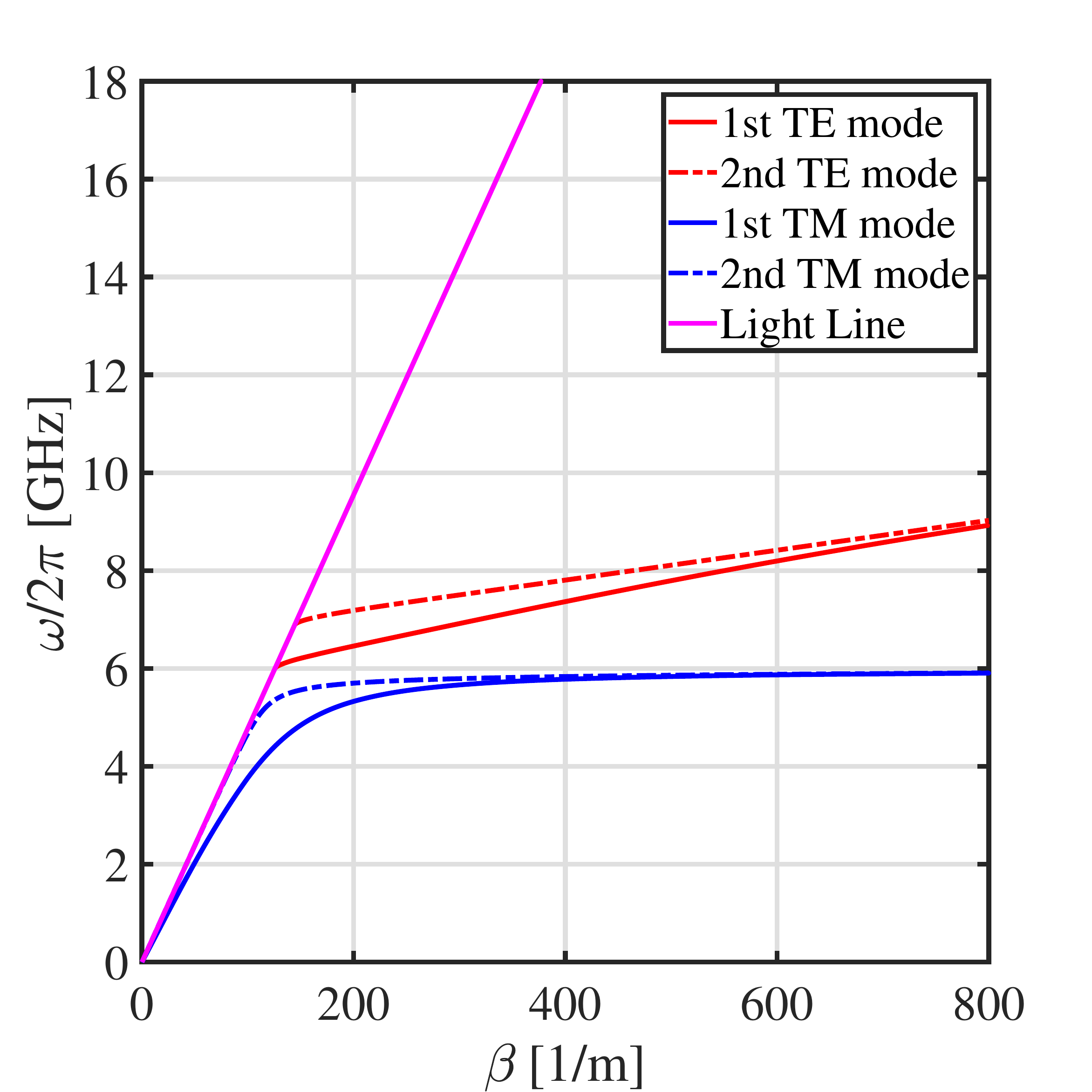}}
	\subfigure[]{\includegraphics[width=0.24\textwidth]{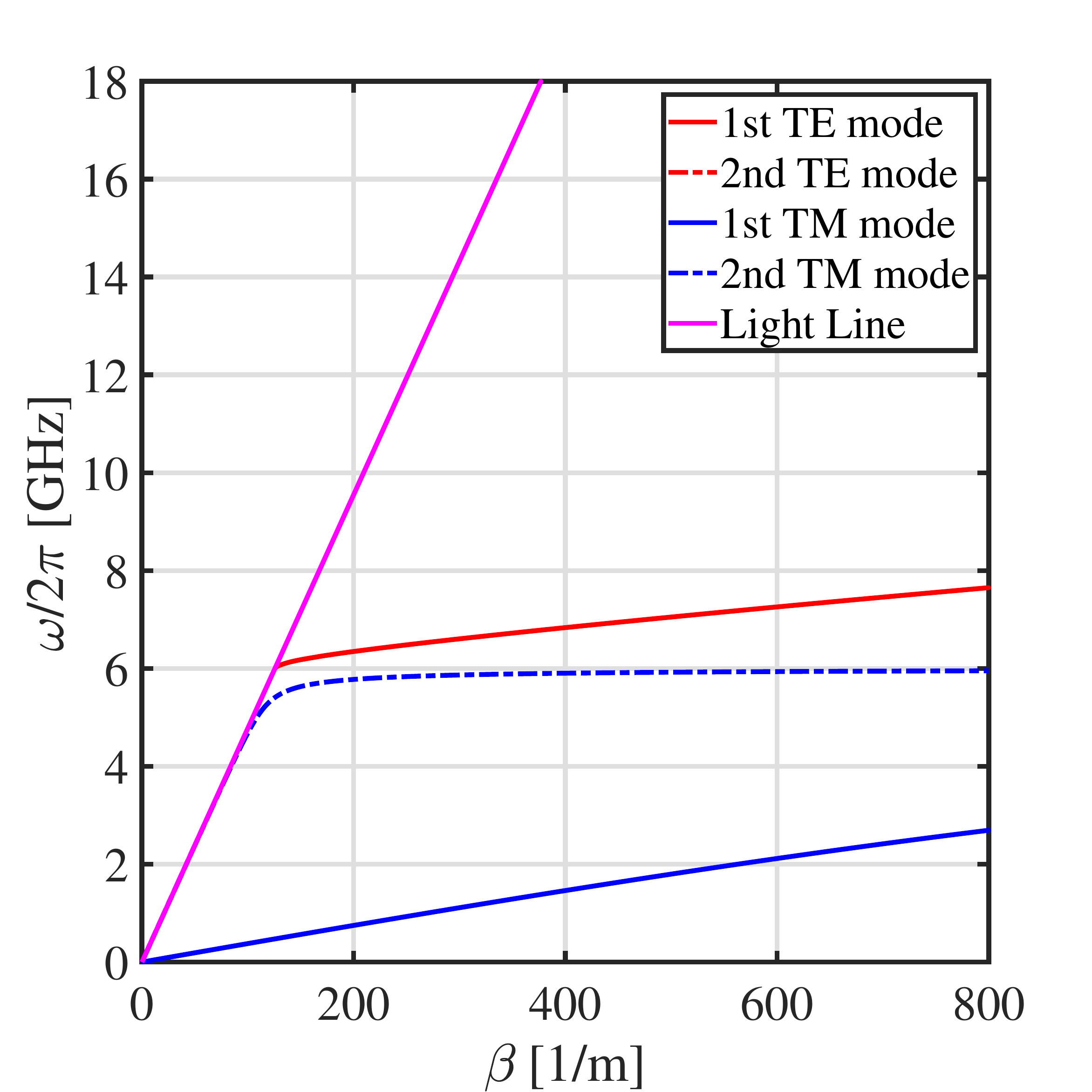}}
	\subfigure[]{\includegraphics[width=0.24\textwidth]{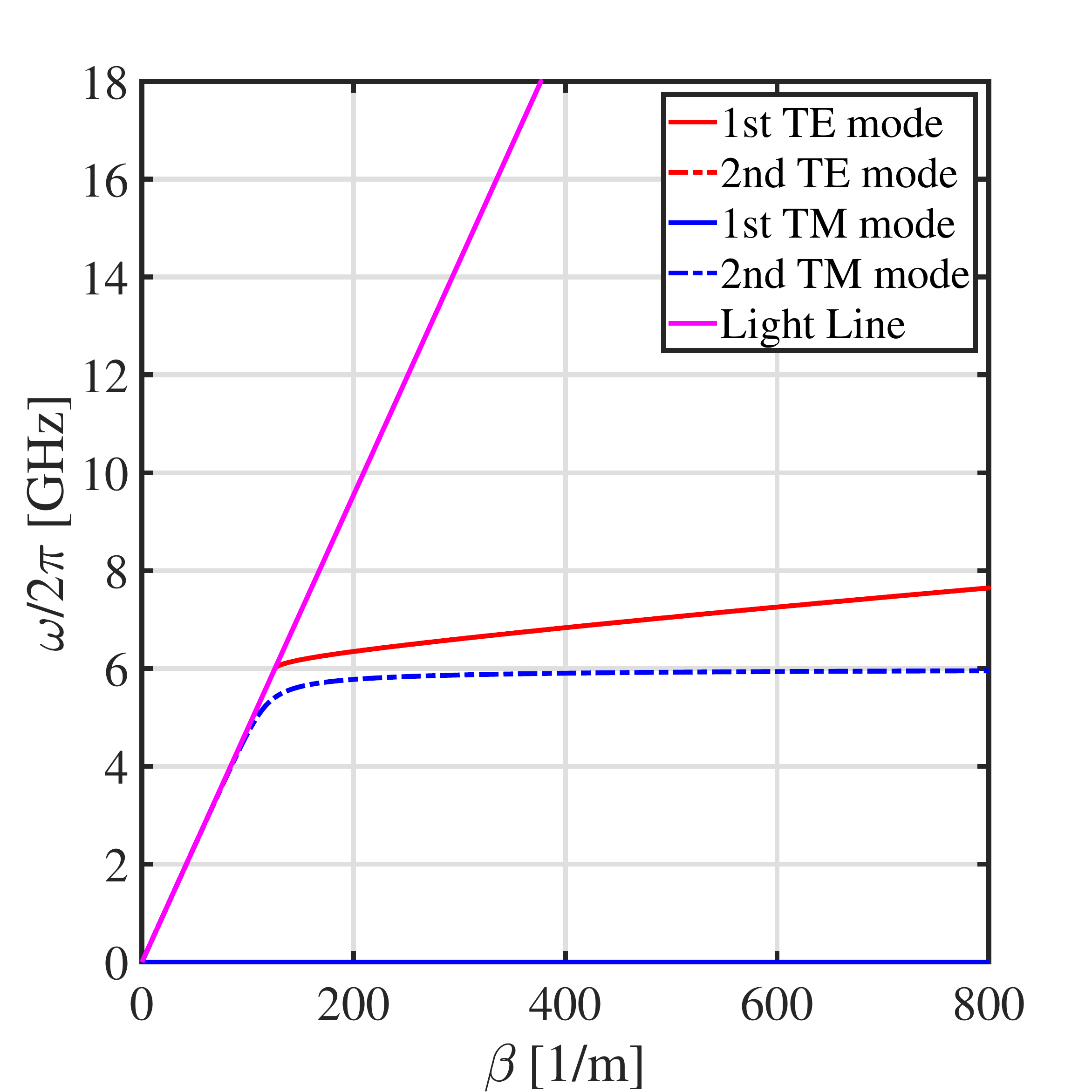}}
	\caption{Dispersion curves of similar impedance sheets for parallel connections in which distances are (a) $d=\lambda_{\rm{6GHz}}$, (b) $d=\lambda_{\rm{6GHz}}/10$, (c) $d=\lambda_{\rm{6GHz}}/5000$, and (d) $d=\lambda_{\rm{6GHz}}/5000000000$.}
	\label{fig:similar parallel}
\end{figure*}

As an example, we assume 6 GHz as the resonance frequency of the symmetric structure, in which $L_2=L_3=1$~nH, $C_2=C_3=0.7$~pF, while $f_2=6$ GHz and $f_3=10$ GHz are the values for the asymmetric structure where $L_2=1$~nH, $C_2=0.7$~pF, $L_3=1$~nH and $C_3=0.25$~pF. 
Figure~\ref{fig:similar parallel} illustrates the dispersion curves of two identical sheets for different distance. When $d=\lambda_{\rm{6GHz}}$, two TM modes are supported below the resonance frequency and two TE modes can propagate above the resonance frequency. Similarly, the two TE (TM) modes have approximately same phase velocity within the symmetric structure and the two TE (or TM) modes separate to each other as the distance decrease. When $d$ approaches  zero, the second TE mode goes to infinity and the first TM mode goes to zero, which gives rise to the propagation of only one TM mode and one TE mode. Interestingly, a very narrow stop band emerges at $f_{\rm{cut-off}}$. Recall that there is no stop band for the symmetric structure in the first scenario.

Similarly to the first scenario, the field is bounded at the symmetric sheets. When the distance between the two sheets is large, the field amplitude is minuscule after a certain distance from the interface in the structure. The fields at two sheets couple with each other when the distance is small. When the two sheets get close to each other, only the in-phase TM mode can propagate along the  $z$-axis. For the TE polarization, only the out-of-phase mode is supported.

\begin{figure*}[t!]
\centering
	\subfigure[]{\includegraphics[width=0.5\columnwidth]{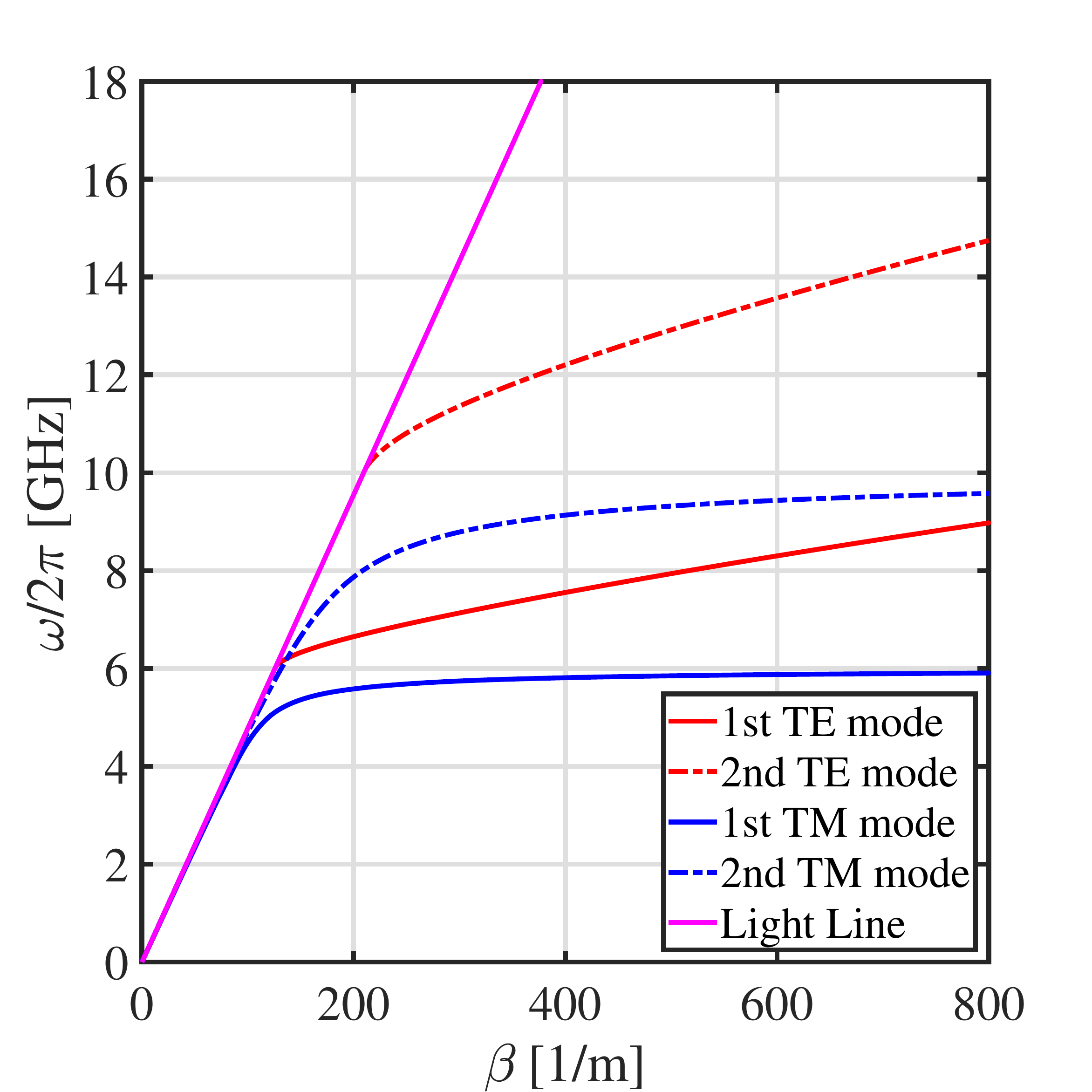}}
	\subfigure[]{\includegraphics[width=0.5\columnwidth]{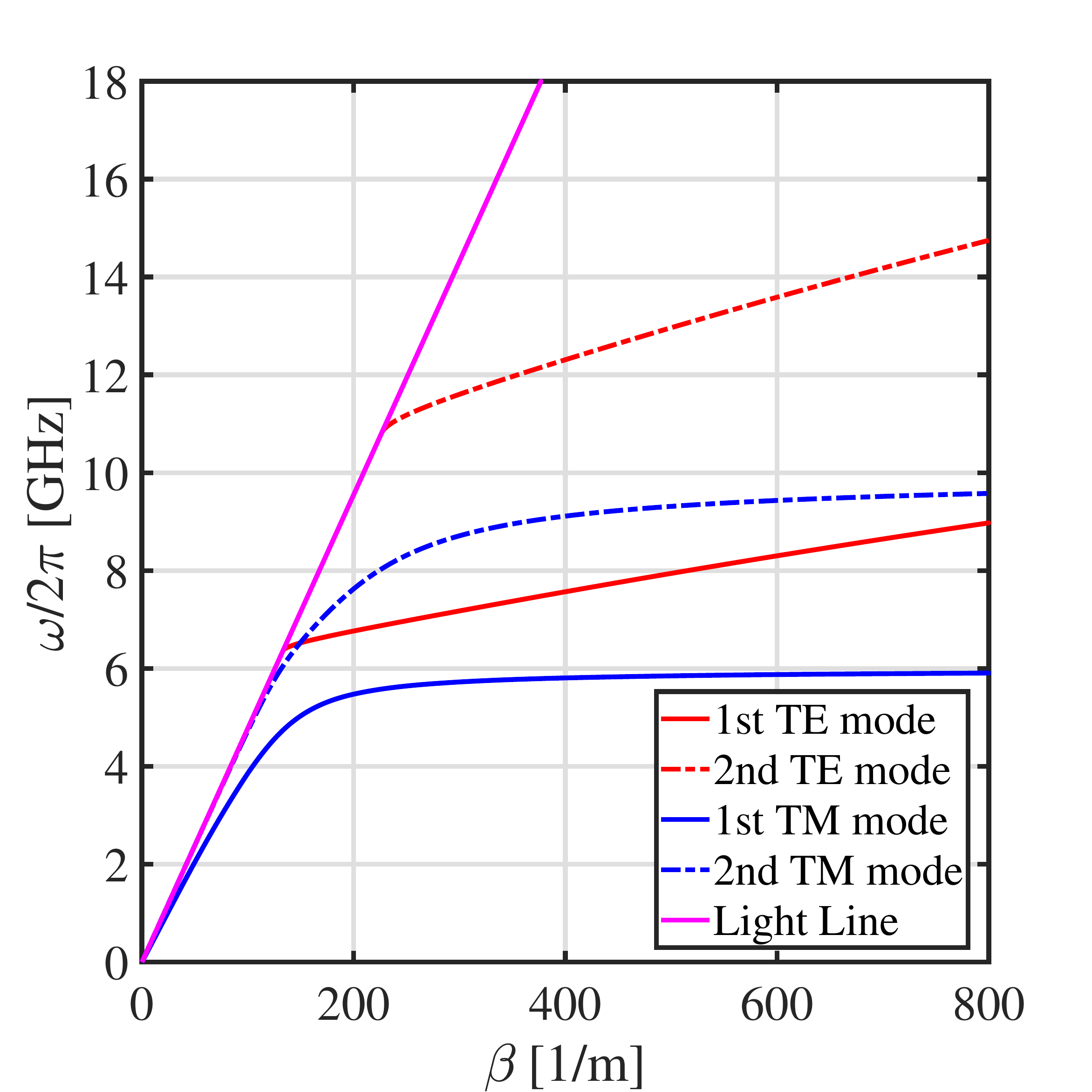}}
	\subfigure[]{\includegraphics[width=0.5\columnwidth]{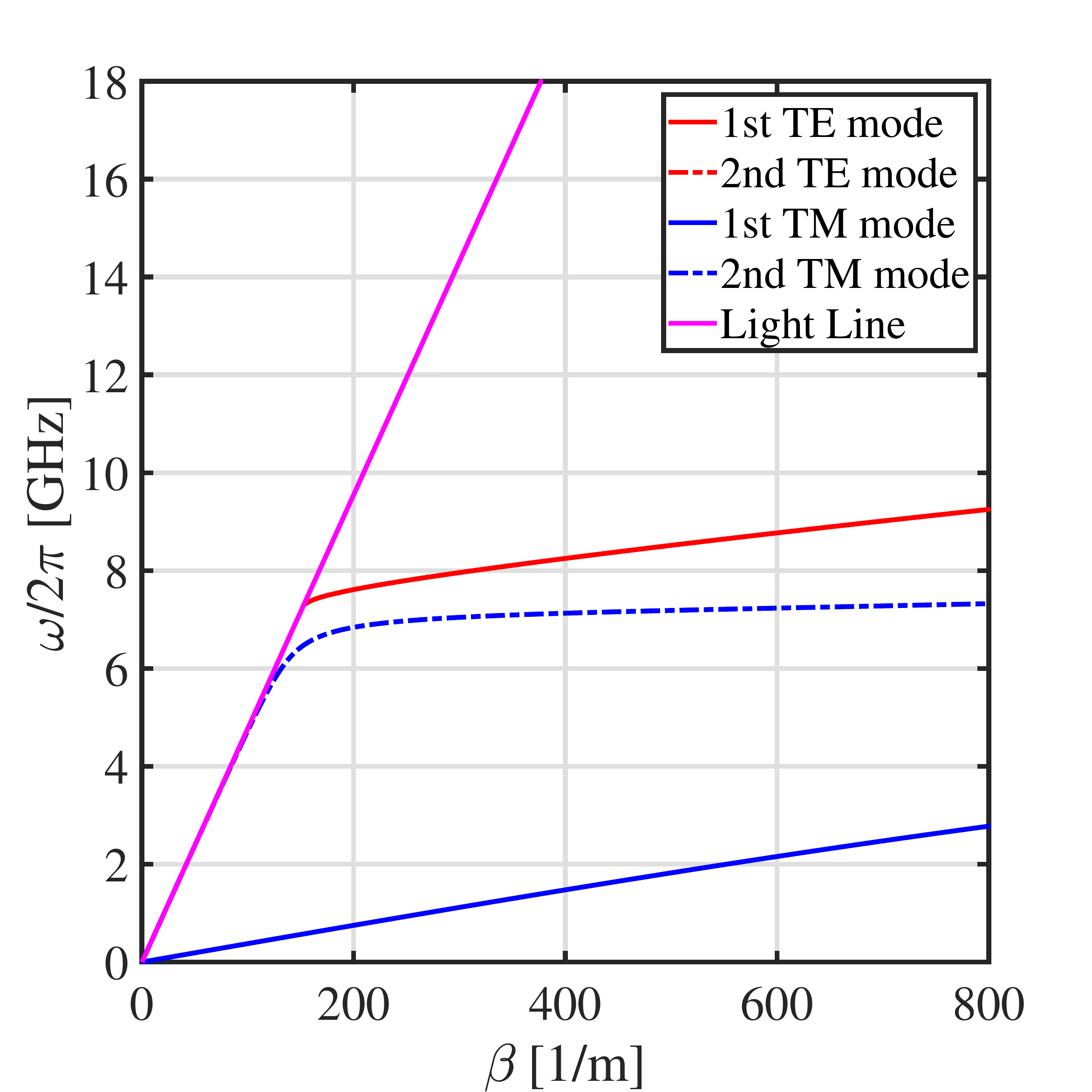}}
	\subfigure[]{\includegraphics[width=0.5\columnwidth]{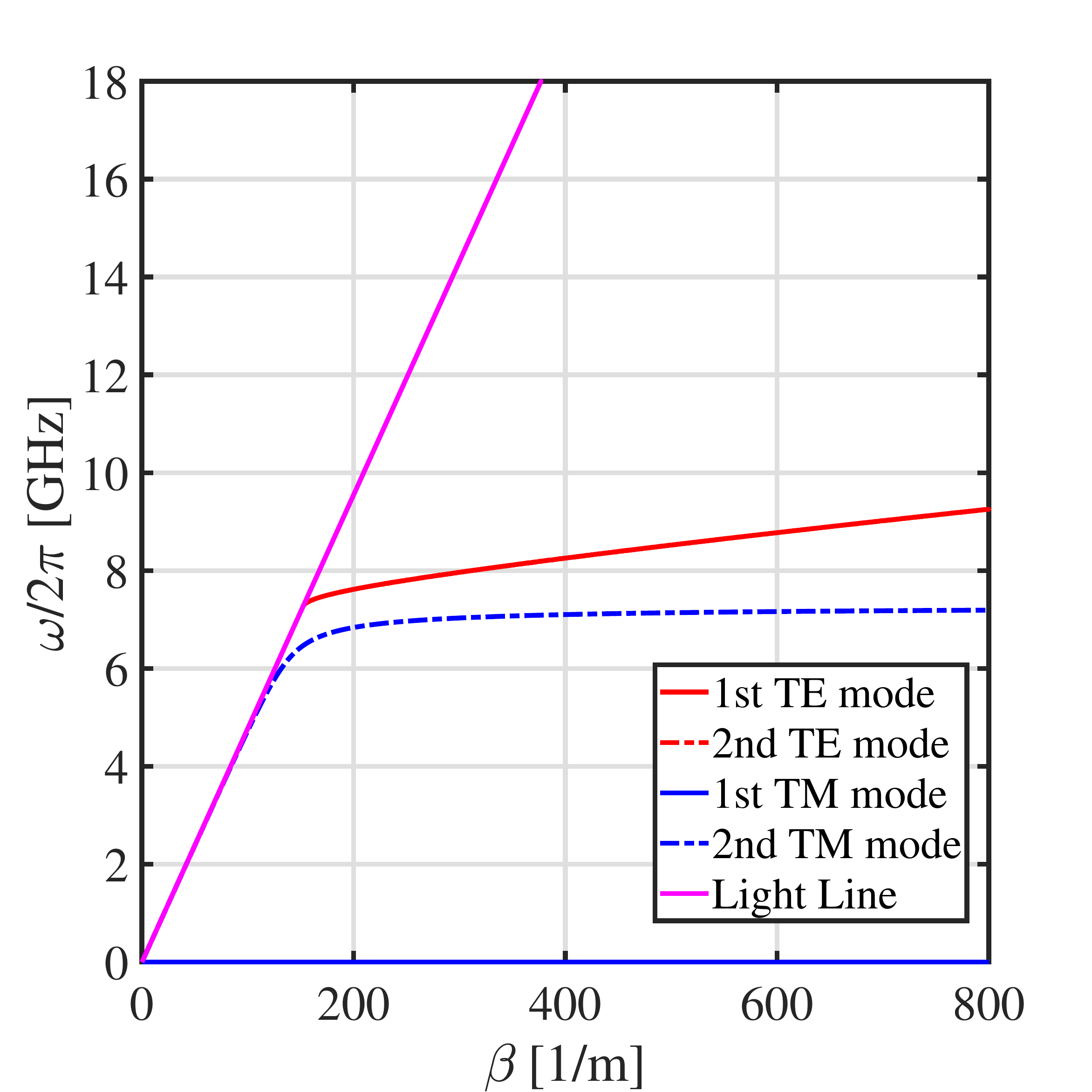}}		
	\caption{Dispersion curves for different impedance sheets for parallel connections in which the distances are (a) $d=\lambda_{\rm{6GHz}}$, (b) $d=\lambda_{\rm{6GHz}}/10$, (c) $d=\lambda_{\rm{6GHz}}/5000$, and (d) $d=\lambda_{\rm{6GHz}}/5000000000$.}
	\label{fig:different parallel}
\end{figure*}

\begin{figure}[t!]\centering 	
     \centerline{\includegraphics[width=0.6\columnwidth]{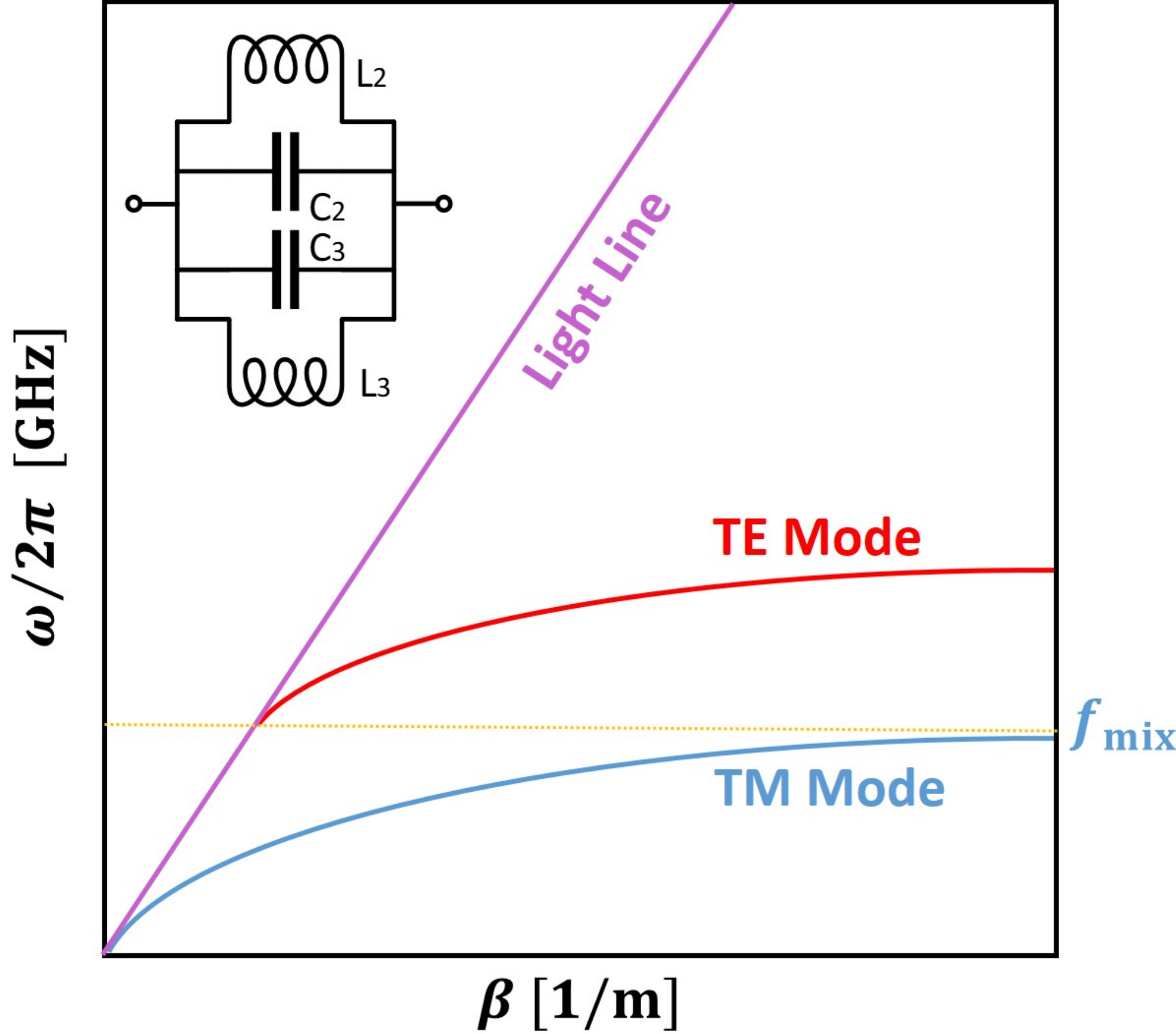}}
	\caption{Dispersion curves for different impedance sheets for parallel connections when the distance $d$ tends to zero.}
	\label{fig:parallel extreme}
\end{figure}

Figure~\ref{fig:different parallel} depicts the dispersion diagram for asymmetric sheets with a parallel connection impedance for different distances. Two TM modes are supported below 6 GHz since both sheets are inductive. In the range from 6 to 10 GHz, waves of both TE and TM polarizations can propagate because one sheet is capacitive and the other one is inductive.  Above 10 GHz, two TE modes can exist because both sheets are capacitive. As the distance decreases, the frequencies of the second TE mode shift towards infinity, and the first TM mode propagates at very small frequencies, which gives rise to a similar behavior with the symmetric structure in the second scenario, but different from the asymmetric case of the series connection, in which four modes survive. When $d$ is approaching zero but still not exactly zero, a new resonance frequency emerges, which  can be found as:
\begin{equation}
f_{\rm{mix}}\approx{{1 \over 2 \pi} \sqrt{L_2+L_3\over L_2 L_3 (C_2+C_3)} }.
\label{eq:fmix parallel}	
\end{equation}
The symmetric structure is a special case of different impedance sheets when $L_2=L_3$ and $C_2=C_3$. It can be discerned from the above equation that  $f_{\rm{mix}}$ of the symmetric structure is equal to $f_2$. The dispersion curves for the extreme case are  illustrated in Fig.~\ref{fig:parallel extreme}. 

%%%%%%%%%%%%%%%%%%%%%%%%%%%%%%%%%%%%%%%%%%%%%%%%%%
\subsection{Third scenario: Sheet impedances as one series and one parallel connections of reactances}
\label{subsec:parallel and series}
In this scenarios, the waveguide under study consists of one series-connection metasurface and one parallel-connection metasurface. We use the notations  $Z_2=j\omega L_2+1/j \omega C_2$ and $Z_3=j\omega L_3/(1-\omega^2 L_3C_3)$. Substituting $Z_2$ and $Z_3$ in  \eqref{eq:gedisrelz3z2} and  \eqref{eq:DisR}, the dispersion equation for TM-modes can be found:
\begin{equation}
\begin{split}
&\Big[2\epsilon_0\alpha C_2 C_3 L_2 L_3+4\epsilon_0^2 C_2 L_2 L_3\Big]\omega^4-\cr
&\Big[2\epsilon_0\alpha(C_2 L_2+ C_3 L_3+C_2 L_3)+\alpha^2\Big(1-e^{-2\alpha d}\Big)C_2C_3 L_3   +\cr
&4\epsilon_0^2L_3\Big]\omega^2+2\epsilon_0\alpha+\alpha^2\Big(1-e^{-2\alpha d}\Big)C_2=0.
\end{split}
\end{equation}
It is worth noting that for this polarization, one of the modes has a cut-off frequency which we denote as $f_2$. The dispersion equation  for the TE polarization reads:
\begin{equation}
\begin{split}
\Big[2\mu_0\alpha C_2C_3L_2L_3+\mu_0^2\Big(1-e^{-2\alpha d}\Big)L_3C_2C_3\Big]\omega^4-\cr
\Big[4\alpha^2C_2L_2L_3+2\mu_0\alpha ( C_2L_2+C_3L_3+C_2L_3)+\cr
\mu_0^2\Big(1-e^{-2\alpha d}\Big)C_2\Big]\omega^2+4\alpha^2 L_3+2\mu_0\alpha=0.
\end{split}
\end{equation}
For this polarization, the modes suffer from cut-off at 
\begin{equation}
f_{\rm{cut-off}}\approx{\frac{b_1+a_3b_2\pm \sqrt{(b_1+a_3b_2)^2-4a_1c_1-4a_2a_3c_1}}{2(a_1+a_2a_3)}},
\end{equation}
where $a_1=2\mu_0C_2C_3L_2L_3$, $a_2=\mu_0^2C_2C_3L_3$, $a_3=2d$, $b_1=2\mu_0(C_2L_2+C_3L_3+C_2L_3)$, $b_2=\mu_0^2C_2$, $c_1=2\mu_0$.

%%%%%%%%%%%%%%%%%%%%%%%%%%%%%%%%%%%%%%%%%%%%%%%%%%%%%%%%%%%%%%%%%%%%%%%%%%%%%%%%%%%%%%%%%%%%%%%%%%%%%
\section{Conclusion}
\label{sec:con}

In this paper, we have introduced a novel waveguiding structure, formed by two penetrable metasurfaces which can have arbitrary impedance properties, for the purpose of guiding the electromagnetic energy and realizing various resonant regimes. We have theoretically studied possible guided modes and derived the dispersion equations. We have shown that in general there are two  propagating modes with transverse magnetic (TM) or transverse electric (TE) polarizations which attenuate outside the structure. The polarizations of those modes depend on the sheet impedances of the metasurfaces. If both metasurfaces are inductive, two modes with TM polarization can propagate along the structure. In the case of two capacitive metasurfaces, the two modes have the TE polarization. Both polarizations can simultaneously contribute to transferring power if one metasurface is inductive and the other one is capacitive. Furthermore, we have investigated the extreme scenarios when the ratio between the surface reactances of the two metasurfaces goes to infinity or approaches zero. We have clearly shown how the dispersion digram of the modes changes as the ratio goes to the extremes. Additionally, we have studied the gain-and-loss waveguide structures in which one metasurface is lossy and the other one is active. In this scenario, both metasurfaces must have equal loss and gain, and the reactive parts of the impedance surfaces must be equal to each other. That is, only parity-time symmetric arrangements support guided waves. Finally, we have considered practical realizations of our proposed waveguides  doing full-wave simulations of  dispersion diagrams. The simulated results are in good agreement with the theoretical results, which confirms our theoretical findings. The found possibilities to realize modes with nearly flat dispersive curves (modes having arbitrary wavenumbers at a given frequency) are promising for applications in near-field superlenses.

\end{document}